\documentclass[aps,preprint,superscriptaddress,nofootinbib]{revtex4-1}
\pdfoutput=1
\usepackage{physics,amsmath,amssymb, hyperref, color,graphicx}
\begin{document}
\count\footins = 1000
\title{Polychronic Tunneling:
New Tunneling Processes Experiencing Euclidean and Lorentzian Evolution Simultaneously
}
\newcommand{\bfx}{{\bf x}}
\newcommand{\bfy}{{\bf y}}
\newcommand{\NO}[1]{\textcolor{blue}{[NO: #1]}}
\newcommand{\my}[1]{\textcolor{magenta}{[MY: #1]}}
\newcommand{\ys}[1]{\textcolor{red}{[YS: #1]}}

\renewcommand{\theequation}{\thesection.\arabic{equation}}

\author{Naritaka Oshita}
\affiliation{RIKEN iTHEMS, Wako, Saitama, Japan, 351-0198}
\author{Yutaro Shoji}
\affiliation{Racah Institute of Physics, Hebrew University of Jerusalem, Jerusalem 91904, Israel}
\author{Masahide Yamaguchi}
\affiliation{Department of Physics, Tokyo Institute of Technology, 2-12-1 Ookayama, Meguro-ku, Tokyo 152-8551, Japan}

\preprint{RIKEN-iTHEMS-Report-21}

\begin{abstract}
    We discuss new possible tunneling processes in the presence of gravity. We formulate quantum tunneling using the Wheeler-deWitt canonical quantization and the WKB approximation. The distinctive feature of our formulation is that it accommodates the coexistence of Euclidean and Lorentzian evolution. It opens up a new possibility of quantum tunneling; {\it e.g.} a bubble wall itself tunnels the potential barrier pulling the field nearby, where the wall region experiences the Euclidean evolution while the other regions experience the Lorentzian evolution simultaneously.
    We execute numerical analysis and find that such a process can have a much higher tunneling rate than that of the Coleman-De Luccia bounce. We also find that the new tunneling processes exist even in the decoupling regime of gravity and affect low energy phenomenology.
\end{abstract}

\maketitle

\section{Introduction}
\setcounter{equation}{0}
Tunneling processes in quantum field theory have played important roles in particle physics and cosmology. One major role is to constrain various new physics models that exhibit instability of the electroweak vacuum. Another role is to discuss phase transitions in various cosmological scenarios, such as inflation \cite{Sato:1980yn,Guth:1980zm}, baryogenesis \cite{Kuzmin:1985mm,Shaposhnikov:1986jp,Shaposhnikov:1987tw}, string landscape \cite{Susskind:2003kw}, and the creation of a universe from nothing \cite{Vilenkin:1982de}.

In the absent of gravity, quantum tunneling has been formulated in \cite{Banks:1973ps,Coleman:1977py,Callan:1977pt} and can be understood from tunneling of a many-body system \cite{Banks:1973ps,Coleman:1977py}, from the imaginary part of the false vacuum energy \cite{Callan:1977pt}, and recently from a more dynamical setup \cite{Andreassen:2016cvx}. Thus, it is fair to say that a consensus has been reached on how the tunneling rate should be calculated;
it is expressed as $\gamma=\mathcal Ae^{-\mathcal B}$,
where $\mathcal B$ is the action of the so-called bounce minus that of the false vacuum, and $\mathcal A$ is an overall factor determined at the loop level.

Even without gravity, there is a subtlety when we treat the so-called mixed tunneling in many-body quantum mechanics. One example of it is a coupled system of two particles where only one of them tunnels. In such a system, it is known that the exponent of the WKB solution is generally complex \cite{PhysRevA.41.32,doi:10.1063/1.466899,doi:10.1063/1.468526}.
However, in quantum field theory, the exponent is either real or pure imaginary at the tree level and complex corrections appear only through perturbative expansions with analytic continuation. Thus, the non-perturbative treatment of the complex exponent seems to be missing in quantum field theory.

Despite its importance, quantum tunneling in the presence of gravity has not been understood adequately.
By analogy with the non-gravitational case, a tunneling formula was first proposed in \cite{Coleman:1980aw} and qualitative features were found reasonable. However, there have been discussions on its rigorousness. One of the issues is that there appear an infinite number of negative modes around the bounce, and the Euclidean action is generally unbounded \cite{Lavrelashvili:1985vn,Tanaka:1992zw,Tanaka:1999pj,Lee:2014uza,Gregory:2018bdt,Bramberger:2019mkv,Jinno:2020zzs}. It spoils the formulation of quantum tunneling relying on the existence of a single negative mode and also obscures what is meant by the bounce with the smallest non-trivial action.

There is also a subtlety about the concept of energy in quantum gravity. Aside from total derivative terms, the Hamiltonian of quantum gravity is given by a sum of constraints and needs to be zero.
Thus, we cannot simply rely on energy to formulate quantum tunneling although it is a crucial quantity that determines the behavior of wave functions in quantum mechanics. 
In addition, if we formulate it with the square root of ``the total energy,'' as in quantum mechanics, it becomes ambiguous whether we should include the space outside of a causal patch.

Furthermore, there is an ambiguity of what should be subtracted from the bounce action to obtain $\mathcal B$. In the absence of gravity, there is only a possibility, which is the action of the false vacuum. It originates from the energy eigenvalue of a time-independent Shr\"odinger equation, or from the dilute gas approximation for multi-bounce configurations. However, in the presence of gravity, we cannot rely on such formulations due to the reasons explained above. In fact, we will see that $\mathcal B$ is not determined solely by a classical solution but also depends on the direction of the field deformation in our formulation.

These subtleties can be attributed to the incomplete formulation of a tunneling process in quantum gravity. 
A promising way to formulate quantum tunneling would be to rely on the Wheeler-deWitt (WdW) canonical quantization \cite{DeWitt:1967yk,Wheeler:1968iap} of quantum gravity.
Although the formulations of quantum tunneling in this direction have been discussed over the decades \cite{Tanaka:1992zw,Gen:1999gi,Mukohyama:1997gb,Kristiano:2018oyv,Cespedes:2020xpn,Garcia-Compean:2021syl}, a rigid formulation for a general deformation path has been lacking.

In this paper, we formulate quantum tunneling using the WdW canonical quantization and the WKB approximation. Our formula possesses locality and tunneling rates can be evaluated independently of the space outside of what is in consideration. In addition, the decaying factor of a WKB wave functional gives the absolute tunneling rate directly and thus we do not need an ad-hoc normalization of the rate. Our formulation is applicable to arbitrary deformation paths and even accommodates the coexistence of Euclidean and Lorentzian evolution, which we call the polychronic evolution.

We confirm the consistency of our formula by showing that all the four-dimensional equations of motion are reproduced from the Hamiltonian constraint, the momentum constraints and the stationary conditions.
We also examine two well-studied examples: quantum tunneling in the mini-superspace and the Coleman-De Luccia (CDL) bubble nucleation. We find they have different tunneling rates even with the same classical solution, which is supported by the discussions in \cite{Tanaka:1992zw,Gen:1999gi}; for the bubble nucleation in a de Sitter vacuum, $\mathcal B$ is given by the bounce action plus the Bekenstein-Hawking entropy of the cosmological horizon, whereas for the mini-superspace, $\mathcal B$ is the bounce action itself.

The most distinctive feature of our formulation is that it accommodates the polychronic evolution.
It not only enables us to deal with the space outside of our consideration consistently, but also allows us to discuss tunneling processes that have not been considered.
In the CDL formulation, a small virtual bubble appears first and its wall moves outward until the bubble materializes.
In our formulation, on the other hand, we can also consider a process in which, for example, the wall of a larger bubble directly tunnels the potential barrier and it simultaneously pulls the field nearby. It causes the Euclidean evolution around the wall region and the Lorentzian evolution in the other regions, which cannot be described in the CDL formulation. We optimize the field evolution numerically and find that the tunneling rate of such a process is much higher than that of the CDL bounce. We also show that such processes exist even in the decoupling regime of gravity, affecting phenomenology at an energy scale much lower than the Planck scale.

The rest of the paper is organized as follows. In the following section, we present our formulation; we review the WdW canonical quantization, perform the WKB approximation, derive the stationary conditions, and define the tunneling rate. Section \ref{sec_minisuperspace} is devoted to the mini-superspace example, and Section \ref{sec_cdl} is to the CDL example. The new possibility of tunneling processes, the polychronic tunneling, is discussed in Section \ref{sec_wall_tunneling}. Finally, we give a summary and discussions in Section \ref{sec_summary}.

\section{Formulation}\label{sec_formulation}
\setcounter{equation}{0}
\subsection{Wheeler-deWitt canonical quantization}
Here, we show our setup and give a brief review on the Wheeler-deWitt formulation of quantum gravity.

We consider the action given by
\begin{equation}
    S=\int \dd[4]{x}\sqrt{-g}\qty[\frac{1}{2\kappa}\mathcal R-\frac12g^{\mu\nu}\pdv{\phi}{x^\mu}\pdv{\phi}{x^\nu}-V(\phi)],\label{eq_action}
\end{equation}
where $\phi$ is a scalar field, $\mathcal R$ is the Ricci scalar and $\kappa=8\pi G$ with $G$ being Newton's constant. We assume the scalar potential, $V(\phi)$, has a false vacuum at $\phi_{\rm F}$ and a true vacuum at $\phi_{\rm T}$. 
We parametrize the metric as
\begin{equation}
    g_{\mu\nu}\dd{x^\mu} \dd{x^\nu}=(-N^2+N_iN^i)\dd{t}^2+2N_i\dd{x^i}\dd{t}+h_{ij}\dd{x^i}\dd{x^j},
\end{equation}
where $N$ is called the lapse function and $N_i$'s are called the shift functions.
The Arnowitt-Deser-Misner (ADM) Hamiltonian is a linear combination of constraints together with total derivative terms,
\begin{align}
    H&=\lambda\pi_N+\lambda_i\pi^i_N+N\mathcal H+N_i\mathcal H^i+\partial_\mu\mathcal H_{\rm bdy}^\mu,\label{eq_total_hamiltonian}
\end{align}
where $\lambda$ and $\lambda_k$ are Lagrange multipliers and\footnote{In Eq.~\eqref{eq_momentum_const}, we divide $\pi^{ij}$ by $\sqrt{h}$ to remind that $\pi^{ij}$ is a density.}
\begin{align}
    \mathcal H&=\frac{1}{\sqrt{h}}\qty[2\kappa G_{ijkl}\pi^{ij}\pi^{kl}+\frac{1}{2}\pi_\phi^2]+\sqrt{h}\qty[-\frac{1}{2\kappa}{}^{(3)}\mathcal R+\frac{1}{2}h^{ij}(\partial_i\phi)(\partial_j\phi)+V(\phi)],\\
    \mathcal H^i&=(\partial^i\phi)\pi_\phi-2\sqrt{h}\nabla_j\frac{\pi^{ij}}{\sqrt{h}},\label{eq_momentum_const}\\
    \mathcal H_{\rm bdy}^t&=\pi^{ij}h_{ij},\label{eq_h_bdy1}\\
    \mathcal H_{\rm bdy}^i&=2\pi^{ij}N_j-\pi^{kl}h_{kl}N^i+\frac{\sqrt{h}}{\kappa}\partial^iN.\label{eq_h_bdy2}
\end{align}
Here,  ${}^{(3)}\mathcal R$ is the three-dimensional Ricci scalar for $h_{ij}$ and
\begin{equation}
    G_{ijkl}=\frac{1}{2}\qty(h_{ik}h_{jl}+h_{il}h_{jk}-h_{ij}h_{kl}).
\end{equation}
The conjugates of $N$, $N_i$, $\phi$ and $h_{ij}$ are denoted as $\pi_N$, $\pi_N^i$, $\pi_\phi$ and $\pi^{ij}$, respectively. They are explicitly given by
\begin{align}
    \pi_N&=0,~\pi_N^i=0,\\
    \pi^{ij}&=\frac{\sqrt{h}}{2\kappa}\qty(Kh^{ij}-K^{ij}),\label{eq_pi_ij}\\
    \pi_\phi&=\frac{\sqrt{h}}{N}(\partial_t\phi-N^i\partial_i\phi),
\end{align}
where $K=K_{ij}h^{ij}$ and
\begin{equation}
    K_{ij}=\frac{1}{2N}\qty(\nabla_iN_j+\nabla_jN_i-\partial_th_{ij}).\label{eq_K_ij}
\end{equation}
Throughout this paper, we use $\nabla_i$ to denote the three-dimensional covariant derivative.

The primary and the secondary constraints are
\begin{equation}
    \pi_N\approx0,~\pi_N^i\approx0,~\mathcal H\approx0,~\mathcal H^i\approx0,
\end{equation}
which are solved after the canonical quantization of the ADM variables;
\begin{align}
    [\hat h_{ij}(\bfx),\hat \pi^{kl}(\bfy)]&=\frac{i\hbar}{2}\qty(\delta_i^k\delta_j^l+\delta_j^k\delta_i^l)\delta^3(\bfx-\bfy),\\
    [\hat \phi(\bfx),\hat \pi_\phi(\bfy)]&=i\hbar\delta^3(\bfx-\bfy),\\
    [\hat N(\bfx),\hat \pi_N(\bfy)]&=i\hbar\delta^3(\bfx-\bfy),\\
    [\hat N_i(\bfx),\hat \pi_N^j(\bfy)]&=i\hbar\delta_i^j\delta^3(\bfx-\bfy).
\end{align}

We consider their representation on the functional space of $(N,N_i,h_{ij},\phi)$ and identify
\begin{equation}
    \hat \pi^{ij}=-i\hbar\fdv{h_{ij}},~\hat \pi_\phi=-i\hbar\fdv{\phi},~\hat \pi_N=-i\hbar\fdv{N},~\hat \pi_N^i=-i\hbar\fdv{N_i}.
\end{equation}
The constraints are then solved on
\begin{align}
    \Omega=\qty{\eval{\Psi[\phi,h_{ij}]}\hat {\mathcal H}\Psi[\phi,h_{ij}]=0,~\hat {\mathcal H}^k\Psi[\phi,h_{ij}]=0}.
\end{align}
The first equation is called the Hamiltonian constraint, or the Wheeler-deWitt equation. It is written as\footnote{There are ambiguities of operator ordering, but they do not affect our results obtained in the leading-order WKB approximation.}
\begin{equation}
    0=\qty[-\frac{\hbar^2}{2\sqrt{h}}\fdv{\Phi^M(\bfx)}\gamma^{MN}\fdv{\Phi^N(\bfx)}+\sqrt{h}\mathcal V]\Psi,
\end{equation}
where $\Phi^{\phi}=\phi$, $\Phi^{(ij)}=h_{ij}$ and
\begin{align}
    &\gamma^{\phi\phi}=1,~
    \gamma^{\phi (ij)}=0,~
    \gamma^{(ij)(kl)}=4\kappa G_{ijkl},\\
    &\mathcal V=-\frac{1}{2\kappa}{}^{(3)}\mathcal R+\frac{1}{2}h^{ij}(\partial_i\phi)(\partial_j\phi)+V(\phi).\label{eq_def_cal_v}
\end{align}

The second equation is called the momentum constraints and is explicitly given by
\begin{equation}
    0=-i\hbar\qty[(\partial^i\phi)\fdv{\phi(\bfx)}-2\sqrt{h}\nabla_j\frac{1}{\sqrt{h}}\fdv{h_{ij}(\bfx)}]\Psi.\label{eq_mom_const}
\end{equation}
Expanding the covariant derivative, it is rewritten in a more useful expression as\footnote{
With infinitesimal parameters, $\boldsymbol\epsilon$, we have
\begin{equation}
    \Psi[\Phi(\bfx+\boldsymbol\epsilon)]=\Psi[\Phi(\bfx)]+\int\dd[3]{x}\partial_k\qty(2\epsilon_l\fdv{\Psi}{h_{kl}(\bfx)})+\mathcal O(\boldsymbol\epsilon^2).
\end{equation}
Thus, the momentum constraint implies that $\Psi$ is invariant under $\Phi^M(\bfx)\to\Phi^M(\bfx+\boldsymbol\epsilon)$ for the system without boundary.}
\begin{equation}
    \qty[\partial_i\Phi^M(\bfx)]\fdv{\Psi}{\Phi^M(\bfx)}=2\partial_kh_{il}\fdv{\Psi}{h_{kl}(\bfx)}.
\end{equation}

\subsection{WKB approximation}\label{subsec_wkb}
Quantum tunneling is a non-perturbative process and a non-trivial semi-classical solution plays an important role. In the following, we obtain a WKB wave functional at the leading order in $\hbar$.

Before going into details, we summarize the procedure for solving the WdW equation and the momentum constraints.
First, we construct a WKB Ansatz of solutions that is consistent with the structure of the WdW equation. The Ansatz contains only one arbitrary function, $\alpha(s,\bfx)$, and is defined for a given path. Here, the path is a curve in the space of $\Phi^M(\bfx)$. If paths are different, the solutions are different in general.
Second, we substitute the Ansatz into the WdW equation and determine $\alpha(s,\bfx)$ so that it solves the WdW equation.
Third, we try to solve the momentum constraints, but we find that we cannot generally solve them only with $\alpha(s,\bfx)$. This is simply because the number of parameters in the Ansatz is not enough. There are two options; (i) constrain the space of solutions so that the momentum constraints are solved automatically, and (ii) add more parameters into the Ansatz. The former is adopted in this section and the latter is given in Appendix \ref{apx_shift}.
Lastly, we substitute $\alpha(s,\bfx)$ into the Ansatz to obtain a solution. We note that we construct a solution in the vicinity of a path, but do not extend it into the entire space of $\Phi^M(\bfx)$.

We begin by constructing a WKB Ansatz.
We expand $\Psi$ with respect to $\hbar$ as
\begin{equation}
    \Psi[\Phi]=\exp\qty[\frac{i}{\hbar}\Theta^{(0)}[\Phi]+\Theta^{(1)}[\Phi]+\dots].
\end{equation}
Then, we consider an arbitrary path, $\Phi^M(s,\bfx)$, which is a map from real number $s\in[s_i,s_f]$ onto the space of $\Phi^M(\bfx)$. We denote
\begin{align}
    \Psi(s)&=\Psi[\Phi^M(s,\bfx)],\\
    \Theta^{(0)}(s)&=\Theta^{(0)}[\Phi^M(s,\bfx)].
\end{align}
At the leading order in $\hbar$, the WdW equation becomes the so-called Einstein-Hamilton-Jacobi equation,
\begin{equation}
    0=\frac{1}{2\sqrt{h}}\gamma^{MN}\fdv{\Theta^{(0)}(s)}{\Phi^M(\bfx)}\fdv{\Theta^{(0)}(s)}{\Phi^N(\bfx)}+\sqrt{h}\mathcal V,\label{eq_ehj}
\end{equation}
where
\begin{equation}
    \fdv{\Theta^{(0)}(s)}{\Phi^M(\bfx)}=\eval{\fdv{\Theta^{(0)}[\Phi^M(\bfx)]}{\Phi^M(\bfx)}}_{\Phi^M(\bfx)=\Phi^M(s,\bfx)}.
\end{equation}

To construct a WKB solution along the path, the tangent vector of the path needs to be proportional to $v^M(s,\bfx)$ defined as
\begin{equation}
    v^M(s,\bfx)=\frac{\alpha(s,\bfx)}{\sqrt{h}}\gamma^{MN}\fdv{\Theta^{(0)}(s)}{\Phi^N(\bfx)},
    \label{eq_def_v}
\end{equation}
where $\alpha(s,\bfx)$ is an arbitrary function. Absorbing the coefficient of proportionality in $\alpha(s,\bfx)$, we take
\begin{equation}
    \pdv{\Phi^M(s,\bfx)}{s}=v^M(s,\bfx).\label{eq_prop_v}
\end{equation}
This proportionality is because of the structure of the higher-order WKB formulas. For the next order, we have
\begin{equation}
    v^M(s,\bfx)\fdv{\Theta^{(1)}(s)}{\Phi^N({\bf x})}=-\frac{\alpha(s,\bfx)}{2\sqrt{h}}\qty[\gamma^{MN}\frac{\delta^2{\Theta^{(0)}(s)}}{\delta{\Phi^M({\bf x})}\delta{\Phi^N({\bf x})}}+\fdv{\gamma^{MN}}{\Phi^M({\bf x})}\fdv{\Theta^{(0)}(s)}{\Phi^N({\bf x})}].
\end{equation}
It contains only the information about the directional derivative of $\Theta^{(1)}(s)$ in the direction $v^M(s,\bfx)$, and those in the other directions remain arbitrary. The same thing happens at the higher orders of WKB. This ambiguity is the ambiguity of the shape of the waves perpendicular to their motion: plane waves, spherical waves and a wave packet have different phase and amplitude dependencies toward such directions.
The constraint of $\pdv{\Phi^M({\bf x})}{s}\propto v^M(s,\bfx)$ makes the path follow the velocity vector of waves and enables unambiguous integration along the path. This can be understood intuitively in the following way. In quantum mechanics, plane waves describe a bunch of particles moving toward the same direction. We are interested in the motion of a single particle, but not in the shape of the bunch since the other particles just came from different initial positions.

Instead of restricting the path, it is more useful to see the above equation as a constraint on $\Theta^{(0)}$ for a given path;
\begin{equation}
    \fdv{\Theta^{(0)}(s)}{\Phi^M(\bfx)}=\frac{\sqrt{h}}{\alpha}\gamma_{MN}\pdv{\Phi^N(s,\bfx)}{s}.\label{eq_cl_path}
\end{equation}
It can be solved along the path as\footnote{If there are degrees of freedom that do not obey the WdW equation, {\it e.g.} those on a physical boundary, one needs to add them at this point explicitly. They can be treated separately and give additional contributions to $\Theta^{(0)}$.}
\begin{align}
    \Theta^{(0)}(s_f)-\Theta^{(0)}(s_i)&=\int_{s_i}^{s_f}\dd{s}\dv{\Theta^{(0)}(s)}{s}\nonumber\\
    &=\int_{s_i}^{s_f}\dd{s}\int\dd[3]{x}\fdv{\Theta^{(0)}(s)}{\Phi^M(\bfx)}\pdv{\Phi^M(s,\bfx)}{s}\nonumber\\
    &=\int_{s_i}^{s_f}\dd{s}\int\dd[3]{x}\frac{2\sqrt{h}}{\alpha}\mathcal K,\label{theta_var_2}
\end{align}
where
\begin{align}
    \mathcal K&=\frac12\gamma_{MN}\pdv{\Phi^M(s,\bfx)}{s}\pdv{\Phi^N(s,\bfx)}{s}\nonumber\\
    &=\frac{(\partial_s\phi)^2}{2}+\frac{G^{ijkl}}{8\kappa}(\partial_sh_{ij})(\partial_sh_{kl}),\label{eq_def_cal_k}
\end{align}
with $\gamma_{MN}$ being the inverse of $\gamma^{MN}$ and
\begin{equation}
    G^{ijkl}=\frac{1}{2}\qty(h^{ik}h^{jl}+h^{il}h^{jk}-2h^{ij}h^{kl}).
\end{equation}
Thus, for a given path, we adopt the WKB Ansatz of\footnote{Obviously, this is just an Ansatz and is still not a solution to the WdW equation without solving $\alpha(s,\bfx)$. This is also true for a similar Ansatz directly derived from Eqs.~\eqref{eq_ehj} and \eqref{eq_cl_path},
\begin{equation}
    \Psi(s_f)=\exp\qty[\frac{i}{\hbar}\int_{s_i}^{s_f}\dd{s}\int\dd[3]{x}2\sqrt{h}\alpha(-\mathcal V)+\order{\hbar^0}]\Psi(s_i).
\end{equation}
The WdW equation is an infinite number of simultaneous differential equations and cannot be solved like the Hamiltonian-Jacobi equation, which is a single differential equation. This point is often confused in literature.}
\begin{equation}
    \Psi(s_f)=\exp\qty[\frac{i}{\hbar}\int_{s_i}^{s_f}\dd{s}\int\dd[3]{x}\frac{2\sqrt{h}}{\alpha}\mathcal K+\order{\hbar^0}]\Psi(s_i),\label{eq_ansatz}
\end{equation}
with Eq.~\eqref{eq_cl_path}. Notice that Eq.~\eqref{eq_cl_path} has additional information about the direction of $\fdv{\Theta^{(0)}(s)}{\Phi^M(\bfx)}$, and thus is superior to Eq.~\eqref{eq_ansatz}.

Let us solve the WdW equation using the Ansatz.
Substituting Eq.~\eqref{eq_cl_path} into Eq.~\eqref{eq_ehj}, we can solve the arbitrary function, $\alpha(s,\bfx)$, as
\begin{equation}
    \alpha(s,\bfx)=\frac{\sqrt{\mathcal K}}{\sqrt{-\mathcal V}}.\label{eq_alpha_def}
\end{equation}
Here, we have chosen the appropriate branch cuts so that the positive direction of $s$ corresponds to the tunneling process\footnote{When $\alpha^2<0$, there appear both a growing solution and a decaying solution. In quantum mechanics, the boundary conditions of the tunneling wave function should be imposed so that the final state has only outgoing waves. Then, the decaying solution quickly dominates over the other inside the potential barrier.
Assuming it is also the case in quantum gravity, we choose the branch cuts so that $\Im\Theta^{(0)}$ becomes positive definite.}. Notice that the path, $\Phi^M(s,\bfx)$, remains arbitrary at this point.

The square of Eq.~\eqref{eq_alpha_def} gives
\begin{align}
 0=C^{\mathcal H}&=\sqrt{h}\qty(\frac{\mathcal K}{\alpha^2}+\mathcal V)\nonumber\\
 &=\sqrt{h}\qty[\frac{(\partial_s\phi)^2}{2\alpha^2}+\frac{G^{ijkl}}{8\kappa\alpha^2}(\partial_sh_{ij})(\partial_sh_{kl})+\frac{h^{ij}}{2}(\partial_i\phi)(\partial_j\phi)+V(\phi)-\frac{{}^{(3)}\mathcal R}{2\kappa}].\label{eq_cl_hamiltonian}
\end{align}
Identifying\footnote{In Appendix \ref{apx_shift}, we obtain the formulas with the shift functions.
Since we do not use the shift functions in the rest of the paper, we continue without them to simplify our discussion.} $N=\alpha$, $N_i=0$ and $t=s$, we see that this is the $(ss)$-component of Einstein's equation, or the classical Hamiltonian constraint.

Next, we examine the momentum constraints. Substituting Eq.~\eqref{eq_cl_path} into Eq.~\eqref{eq_mom_const}, we obtain
\begin{equation}
    0=C^{\mathcal M}_i=\sqrt{h}\qty[\frac{1}{\alpha}(\partial_i\phi)(\partial_s\phi)-\frac{h_{ip}}{2\kappa}\nabla_j\qty(\frac{G^{pjkl}}{\alpha}\partial_s h_{kl})],\label{eq_mom_const2}
\end{equation}
at the leading order in $\hbar$. With the same identification of parameters, these are the $(si)$-components of Einstein's equation, or the classical momentum constraints.
A more useful expression is
\begin{equation}
    0=C^{\mathcal M}_i=\frac{\sqrt{h}}{\alpha}\gamma_{MN}\qty(\partial_i\Phi^M)\qty(\partial_s\Phi^N)-\frac{1}{2\kappa}\partial_k\qty(h_{il}\frac{\sqrt{h}}{\alpha}G^{klpq}\partial_sh_{pq}).
\end{equation}
Now, the path, $\Phi^M(s,\bfx)$, should be taken so that it satisfies the classical momentum constraints\footnote{If we include additional functions, $\beta_i(s,\bfx)$'s, in the Ansatz as in Appendix \ref{apx_shift}, we can solve them instead of restricting $\Phi^M(s,\bfx)$. Then, we can construct an element of $\Omega$ for an arbitrary $\Phi^M(s,\bfx)$ (but except for ones that have points satisfying both $\mathcal K=0$ and $\partial\Phi^M/\partial s\neq0$) in the WKB approximation. The reason why we can construct a solution for an arbitrary path is explained in Subsection \ref{subsec_stationary}.}

Finally, substituting Eq.~\eqref{eq_alpha_def} into Eq.~\eqref{theta_var_2}, we obtain our final result;
\begin{equation}
    \Theta^{(0)}(s_f)-\Theta^{(0)}(s_i)=\int_{s_i}^{s_f} \dd{s}\int\dd[3]{x}2\sqrt{h}\sqrt{-\mathcal V}\sqrt{\mathcal K}.\label{eq_theta_final}
\end{equation}
In the rest of the paper, we take $\Theta^{(0)}(s_i)=0$ to simplify expressions, without the loss of generality. We also omit the argument and denote $\Theta^{(0)}=\Theta^{(0)}(s_f)$.

Notice that the spatial integral is outside of the square root unlike in quantum mechanics or in \cite{Cespedes:2020xpn}. This means that oscillatory or decaying behavior of the wave functional is determined locally, as it should be since the WdW equation is local. This is an important feature since we do not need to worry about the space outside of our consideration; we can separate such regions and consider conditional probabilities consistently.
In Appendix \ref{sec_camq}, we show that our formula agrees with that in \cite{Cespedes:2020xpn} when $\alpha$ does not depend on the spatial position and the signs of $\mathcal V$ and $\mathcal K$ do not flip. However, to calculate a bubble nucleation rate without an ad-hoc normalization of the wave functional, we find that our formulation is indispensable as explained in Appendix \ref{apx_foliation}.

As we have seen in this subsection and will see in the next subsection, $\alpha$ behaves like the lapse function. It is, however, just a function in the Ansatz used to solve the Hamiltonian constraint, and there is nothing that prevents the local flipping of the sign of $\alpha^2$. If we identify $t=s$, $N=\alpha$ and $N_i=0$, the local flipping gives a picture of polychronic evolution, where the four-dimensional metric becomes Euclidean or Lorentzian depending on the spatial regions.

The coexistence of Euclidean and Lorentzian metrics on the same time slice is not a surprising outcome. When we connect the CDL bounce to a Lorentzian solution describing the bubble expansion, there appears a constant-time slice that separates the Euclidean region and the Lorentzian region. If we take a different foliation, however, these generally coexist in the same time slice. In addition, there is a phenomenon called the mixed tunneling in many-body quantum mechanics, where Euclidean and Lorentzian evolution occur simultaneously.

Before closing this subsection, we elaborate on $\alpha=0$ and $|\alpha|=\infty$, where the four-dimensional metric cannot be defined. As we can see from Eq.~\eqref{eq_alpha_def}, $|\alpha|=\infty$ occurs when $\mathcal V=0$ and $\mathcal K\neq0$. Remembering that we are interested in $\Theta^{(0)}[\Phi^M(\bfx)]$ around $\Phi^M(\bfx)=\Phi^M(s,\bfx)$, we go back to the Einstein-Hamilton-Jacobi equation, Eq.~\eqref{eq_ehj}. When $\mathcal V[\Phi^M(\bfy)]=0$, there exists a trivial solution, $\delta\Theta^{(0)}(s)/\delta\Phi^N(\bfy)=0$ for all $N$. Since we are interested in paths that are continuous and bounded, $\partial\Phi^M(s,\bfy)/\partial s$ does not diverge, and we have $[\partial\Phi^M(s,\bfy)/\partial s]
[\delta\Theta^{(0)}(s)/\delta\Phi^M(\bfy)]=0$ in Eq.~\eqref{theta_var_2}, which is consistent with Eq.~\eqref{eq_theta_final}.
If necessary, one may also add a small imaginary part to $V(\phi)$ and make $\alpha$ analytic everywhere.

On the other hand, we need extra care for $\alpha=0$, which occurs when $\mathcal V\neq0$ and $\mathcal K=0$. Since our Ansatz, Eqs.~\eqref{eq_cl_path} and \eqref{theta_var_2}, is not well defined for $\alpha=0$, we go back to Eq.~\eqref{eq_prop_v}. Since $v^M(s,\bfy)=0$ from Eq.~\eqref{eq_def_v}, we have $\partial\Phi^N(s,\bfy)/\partial s=0$ for all $N$. Assuming $\Theta^{(0)}(s)$ is regular around the path, we have $[\partial\Phi^M(s,\bfy)/\partial s][\delta\Theta^{(0)}(s)/\delta\Phi^M(\bfy)]=0$ in Eq.~\eqref{theta_var_2}, and the result is again consistent with Eq.~\eqref{eq_theta_final}. This is reasonable because it says that, if $\Phi^M(s+\delta s,\bfy)=\Phi^M(s,\bfy)$ with infinitesimal variable $\delta s$, there is no contribution to $\Theta(s+\delta s)-\Theta(s)$ from point $\bf y$.
Notice that, since $\gamma^{MN}$ is an indefinite metric, a general path could have $\Phi^M(s+\delta s,\bfy)\neq\Phi^M(s,\bfy)$ with keeping $\mathcal K=0$. Since this is not consistent with our Ansatz, we need to choose paths so that this does not happen.

\subsection{Stationary condition}
\label{subsec_stationary}
In the previous subsection, a WKB solution is obtained for an arbitrary path satisfying the momentum constraints.
Let us first explain why we can take nearly arbitrary paths, which do not necessarily satisfy the equations of motion. This is because no boundary conditions have been assumed in the definition of $\Omega$. For example, we may even take a boundary that closely encloses the path just like a waveguide. Then, we can trivially take arbitrary paths. If necessary, one can also extend\footnote{There is no need for the wave functional to be finite everywhere, which is explained in the beginning of Subsection \ref{subsec_tunneling}.} the solution outside the waveguide by solving the WdW equation outward with the same boundary conditions.

In most cases, however, we do not assume such artificial boundary conditions and there is a natural direction of propagation. We first see the Huygens-Fresnel principle determines the direction as discussed in \cite{Gerlach:1969ph}. Let us assume there is a wave front around $\bar\Phi_0^M(\bfx)$,
\begin{equation}
    \Psi[\bar\Phi_0^M(\bfx)]\simeq\Psi[\bar\Phi_0^M(\bfx)+\lambda^M(\bfx)],
\end{equation}
where $\lambda^M(\bfx)\in \Xi_0$ with $\Xi_0$ being a set of deviations that keep $\Psi$ nearly constant. Let us discuss the value of the wave functional at $\Phi_1^M(\bfx)$, which is close to $\bar\Phi_0^M(\bfx)$. 
For each $\lambda^M(\bfx)$, we choose a linear path as
\begin{equation}
    \Phi^M(s,\bfx)=(\Phi_1^M(\bfx)-\bar\Phi_0^M(\bfx)-\lambda^M(\bfx))s+\bar\Phi_0^M(\bfx)+\lambda^M(\bfx).
\end{equation}
and define
\begin{equation}
    \Theta^{(0)}[\lambda]\simeq\Theta^{(0)}(s=1).
\end{equation}
Then, the Huygens-Fresnel principle states that
\begin{equation}
    \Psi[\Phi_1^M(\bfx)]\simeq\int\dd{\lambda}e^{\frac{i}{\hbar}\Theta^{(0)}[\lambda]}\Psi[\bar\Phi_0^M(\bfx)+\lambda^M(\bfx)].
\end{equation}
It can be approximated by the value at the saddle point as
\begin{equation}
    \Psi[\Phi_1^M(\bfx)]\simeq e^{\frac{i}{\hbar}\Theta^{(0)}[\lambda_*]}\Psi[\bar\Phi_0^M(\bfx)+\lambda_*^M(\bfx)],
\end{equation}
where $\lambda_*^M(\bfx)$ is the saddle point.
To obtain the same result with a single path, the path should pass through $\bar\Phi_0^M(\bfx)+\lambda_*^M(\bfx)$. We choose $\Phi_1^M(\bfx)$ so that $\lambda_*^M(\bfx)=0$, and we denote it as $\bar\Phi_1^M(\bfx)$.
We expect that $\Psi$ around $\bar\Phi_1^M(\bfx)$ is well described by the single path. Then, because of Eq.~\eqref{eq_prop_v}, $\Psi$ is almost constant around $\bar\Phi_1^M(\bfx)$ toward the directions perpendicular to $\partial\Phi^M(s,\bfx)/\partial s$.
Thus, we can repeat the same procedure with changing $\bar\Phi_0^M(\bfx)\to\bar\Phi_1^M(\bfx)$ and obtain a series of $[\bar\Phi_0^M(\bfx),\bar\Phi_1^M(\bfx),\bar\Phi_2^M(\bfx),\cdots,\bar\Phi_n^M(\bfx)]$. If we take $n\to\infty$ with keeping $\bar\Phi_0^M(\bfx)$ and $\bar\Phi_n^M(\bfx)$ constant, the series defines a path that is a saddle point of $\Theta(s)$; $\Phi_{i-1}^M(\bfx)$ is the ``closest'' (or ``farthest'') point on $\Xi_{i-1}$ measured from $\Phi_i^M(\bfx)$.
Although the above explanation is not rigid\footnote{In \cite{Shoji:2022rke}, the path integral formulation is introduced for the WdW equation and the saddle point approximation is found to be reasonable.}, it seems motivating to consider saddle points for generic boundary conditions.
In the following, we derive the stationary conditions and show that they actually reproduce the equations of motion.

For the sake of convenience, we define the four-dimensional metric with $N=\alpha=\frac{\sqrt{\mathcal K}}{\sqrt{-\mathcal V}}$ and $N_i=0$, {\it i.e.}
\begin{equation}
    g_{\mu\nu}\dd{x^\mu} \dd{x^\nu}=-\alpha^2\dd{s}^2+h_{ij}\dd{x}^i\dd{x}^j.
\end{equation}
Notice that $\alpha$ can be real or pure imaginary depending locally on the signs of $\mathcal V$ and $\mathcal K$, and that there is no need for the existence of a corresponding four-dimensional manifold. Since $\alpha=0$ simply means $\partial\Phi^M(s,\bfx)/\partial s=0$, we assume $\alpha\neq0$ in the following.
Hereafter, $K_{ij}$, $K$, $\mathcal R_{ij}$ and $\mathcal R$ are understood as those with $N=\alpha$ and $N_i=0$.

First, taking the variation with respect to $\phi$, we obtain\footnote{Precisely speaking, the variation should be taken under the momentum constraints. In \cite{Shoji:2022rke}, it is checked that it gives the same results.}
\begin{align}
    \delta \Theta^{(0)}&=\int\dd[4]{x}\sqrt{h}\qty[-\frac{\sqrt{\mathcal K}}{\sqrt{-\mathcal V}}\delta\mathcal V+\frac{\sqrt{-\mathcal V}}{\sqrt{\mathcal K}}\delta\mathcal K]\nonumber\\
    &=-\int\dd[4]{x}\alpha\sqrt{h}\qty[\dv{V}{\phi}\delta\phi+h^{ij}(\partial_i\phi)(\partial_j\delta\phi)+g^{ss}(\partial_s\phi)(\partial_s\delta\phi)]\nonumber\\
    &=-\int\dd[4]{x}\alpha\sqrt{h}\qty[\dv{V}{\phi}-D_\mu(\partial^\mu\phi)]\delta\phi.
\end{align}
Here, $D_\mu$ is the four-dimensional covariant derivative and $\dd[4]{x}$ is a short-hand for $\dd{s}\dd[3]{x}$.
The terms inside the parentheses need to be zero, which gives the equations of motion for the scalar field.

Second, taking the variation with respect to $h_{ij}$, we obtain
\begin{align}
    \delta \Theta^{(0)}&=\int\dd[4]{x}\sqrt{h}\qty[\frac{\delta h}{h}\sqrt{-\mathcal V}\sqrt{\mathcal K}-\frac{\sqrt{\mathcal K}}{\sqrt{-\mathcal V}}\delta\mathcal V+\frac{\sqrt{-\mathcal V}}{\sqrt{\mathcal K}}\delta\mathcal K]\nonumber\\
    &=-\int\dd[4]{x}\alpha\sqrt{h}\qty[\mathcal V h^{ij}\delta h_{ij}+\frac{1}{2}(\partial_i\phi)(\partial_j\phi)\delta h^{ij}-\frac{1}{2\kappa}\delta\qty({}^{(3)}\mathcal R+K_{ij}K^{ij}-K^2)_\alpha]\nonumber\\
    &=-\int\dd[4]{x}\alpha\sqrt{h}\qty[\mathcal V h^{ij}\delta h_{ij}-\frac{1}{2}(\partial^i\phi)(\partial^j\phi)\delta h_{ij}-\frac{1}{2\kappa}\delta\qty(\mathcal R+D_\rho \mathcal X^\rho)_\alpha]\nonumber\\
    &=-\int\dd[4]{x}\alpha\sqrt{h}\qty[\mathcal V h^{ij}\delta h_{ij}-\frac{1}{2}(\partial^i\phi)(\partial^j\phi)\delta h_{ij}-\frac{1}{2\kappa}\delta\qty(\mathcal R)_\alpha-\frac{\sqrt{h}}{2\kappa}\delta\qty(\frac{1}{\sqrt{h}})D_\rho\mathcal X^\rho]\nonumber\\
    &=-\int\dd[4]{x}\alpha\sqrt{h}\qty[\mathcal V h^{ij}-\frac{1}{2}(\partial^i\phi)(\partial^j\phi)+\frac{1}{2\kappa}\qty(\mathcal R^{ij}+\frac{D_\rho \mathcal X^\rho}{2}h^{ij})]\delta h_{ij},
\end{align}
where $\mathcal R_{ij}$ is the four-dimensional Ricci tensor, $\delta(\dots)_\alpha$ is the variation with fixing $\alpha$. We have used the Gauss-Codazzi equation;
\begin{align}
    \mathcal R={}^{(3)}\mathcal R+K_{ij}K^{ij}-K^2-D_\rho \mathcal X^\rho,
\end{align}
where
\begin{equation}
    \mathcal X^s=g^{ij}\partial^sg_{ij},~\mathcal X^i=g^{ss}\partial^ig_{ss}.
\end{equation}
Thus, the stationary conditions for $h_{ij}$ are
\begin{align}
    0&=\qty[V(\phi)+\frac{1}{2}h^{kl}(\partial_k\phi)(\partial_l\phi)]h_{ij}-\frac{1}{2}(\partial_i\phi)(\partial_j\phi)+\frac{1}{2\kappa}\qty[\mathcal R_{ij}-\qty({}^{(3)}\mathcal R-\frac{D_\rho \mathcal X^\rho}{2})h_{ij}].\label{eom_d}
\end{align}
The classical Hamiltonian constraint, Eq.~\eqref{eq_cl_hamiltonian}, can be rewritten as
\begin{equation}
    {}^{(3)}\mathcal R-\frac{D_\rho \mathcal X^\rho}{2}=\frac{\mathcal R}{2}+\kappa\qty[V(\phi)+\frac12g^{\mu\nu}(\partial_\mu\phi)(\partial_\nu\phi)]-\kappa g^{ss}(\partial_s\phi)^2.
\end{equation}
Substituting this into Eq.~\eqref{eom_d}, we obtain
\begin{align}
    \mathcal R_{ij}-\frac{\mathcal R}{2}g_{ij}&=\kappa(\partial_i\phi)(\partial_j\phi)-\kappa  g_{ij}\qty[V(\phi)+\frac12g^{\mu\nu}(\partial_\mu\phi)(\partial_\nu\phi)],\label{eq_eom_hij}
\end{align}
which are the $(ij)$-elements of Einstein's equation.

As we have mentioned in the previous subsection, the classical Hamiltonian constraint and the classical momentum constraints correspond to the $(ss)$-component and $(si)$-components of Einstein's equation, respectively. It can be easily checked by using
\begin{align}
    \mathcal R_{ss}-\frac{\mathcal R}{2}g_{ss}&=\frac{\alpha^2}{2}{}^{(3)}\mathcal R-\frac18G^{ijkl}(\partial_sh_{ij})(\partial_sh_{kl}),\\
    \mathcal R_{si}-\frac{\mathcal R}{2}g_{si}&=\frac{\alpha}{2} h_{ik}\nabla_l\qty(\frac{G^{klpq}}{\alpha}\partial_sh_{pq}).
\end{align}
Thus, promoting Eq.~\eqref{eq_eom_hij} to the four-dimensional field equations, we successfully reproduce the full Einstein's equation;
\begin{align}
    \mathcal R_{\mu\nu}-\frac{\mathcal R}{2}g_{\mu\nu}&=\kappa T_{\mu\nu},\label{eq_einstein}
\end{align}
where
\begin{equation}
    T_{\mu\nu}=(\partial_\mu\phi)(\partial_\nu\phi)-g_{\mu\nu}\qty[\frac12g^{\delta\lambda}(\partial_\delta\phi)(\partial_\lambda\phi)+V(\phi)].
\end{equation}
We emphasize that each component has its own meaning; the $(ij)$-components of Einstein's equation give the stationary conditions for $h_{ij}$, while the $(ss)$- and the $(si)$-components are the Hamiltonian and the momentum constraints along the path, respectively. Notice that the stationary conditions are not mandatory for the use of the WKB Ansatz.

In this paper, we use the following combinations for the stationary conditions;\footnote{The first equation is 
\begin{equation}
    \mathcal R_{ij}=\kappa\qty[T_{ij}-h_{ij}\frac{g^{\mu\nu}T_{\mu\nu}}{2}].
\end{equation}
}
\begin{align}
    0&=F_{ij}= \mathcal R_{ij}-\kappa\qty[(\partial_i\phi)(\partial_j\phi)+h_{ij}V(\phi)],\\
    0&=F_\phi=-D_\mu\partial^\mu\phi+\dv{V}{\phi},\label{eq_scalar_eom}
\end{align}
where
\begin{align}
    \mathcal R_{ij}={}^{(3)}\mathcal R_{ij}+\frac{1}{2\alpha}\partial_s\frac{1}{\alpha}\partial_sh_{ij}-\frac{1}{\alpha}\nabla_i\nabla_j\alpha+\frac{h^{kl}}{4\alpha^2}\qty[(\partial_sh_{kl})(\partial_sh_{ij})-2(\partial_sh_{il})(\partial_sh_{kj})],
\end{align}
with ${}^{(3)}\mathcal R_{ij}$ being the three-dimensional Ricci tensor. Notice that the Hamiltonian constraint, the momentum constraints and the stationary conditions are not independent.

\subsection{Tunneling rate}\label{subsec_tunneling}
\subsubsection{Difficulties in Defining Tunneling Rates}
So far, we have discussed the WKB approximation without picking up a specific wave functional in $\Omega$.
To discuss a tunneling process, we need to clarify which wave functional is responsible and how we evaluate the tunneling rate with it. However, it requires an understanding of quantum tunneling beyond the WKB approximation, which poses several difficulties:
\begin{itemize}
    \item Since the WdW equation does not have a time variable, its solution has to be stationary. If the solution is well-defined over all the configuration space of $\Phi^M(\bfx)$, the probability localizes at the true vacuum and there is no flow of probability, which does not seem to describe the tunneling process. One resolution would be to introduce sources and sinks, where the probability has non-zero divergence and the solution becomes singular. Some of sources and sinks can be replaced with the point at infinity in the configuration space. Then, we would obtain stationary solutions that have flows from the sources to the sinks.
    \item Even if we introduce sources and sinks, we need to make sure that all the probability coming from the sources flows into the sinks quickly enough. Otherwise, the probability may accumulate at a point and it may finally flow back disturbing the tunneling wave functional.
    \item In order to calculate the tunneling rate, we need the amplitude of the waves hitting the potential barrier (incoming waves). However, there always exist reflected waves and they interfere with each other everywhere since the wave functional is stationary. It is not an easy task to separate out only the incoming waves from a stationary state even in a multi-dimensional quantum mechanics.
    \item There are difficulties in defining incoming waves and reflected waves. It is not only because there is no time from the beginning, but also because $\mathcal V$ changes over the configuration space. The incoming and the reflected waves easily mix with each other and the definition is not unique. In relation to this, it is also difficult to define asymptotically free waves in the configuration space.
    \item A tunneling rate is originally defined as the probability to observe a tunneling event in unit volume and unit time. However, this is not trivial since there is no global time in our formalism. Although variable $s$ appears in the WKB approximation, we do not have such a variable away from the path.
    \item To determine a tunneling rate uniquely, we need to specify a tunneling wave functional somewhat uniquely. It is, however, not clear what conditions ensure the existence and the uniqueness of a solution. This difficulty comes from the fact that the WdW equation is not a single wave equation but an infinite number of simultaneous wave equations.
    \item The wave functional is not normalizable in general since the configuration space has infinite volume and infinite dimension. Once we go beyond the WKB approximation, the notion of probability suffers from infinity.
    \item The WdW equation itself has not been understood well beyond the WKB approximation. There are operator ordering ambiguities and a regularization problem for operator products.
\end{itemize}

Although it is very tough\footnote{Many of these problems are related to the definition of time. In \cite{Shoji:2022rke}, it is observed that the Hamiltonian constraint describes the time evolution of operators, which would give implications on these time issues.} to define a tunneling rate rigidly, it is fair to say that the decaying factor of the wave functional in the WKB approximation has something to do with the tunneling rate; whatever the full solution is, the amplitudes of incoming waves decay according to the WKB rate along the path. In addition, there is no need to worry about interference with the waves coming from other directions since the interference is ignored at the leading order in the WKB approximation.

\subsubsection{Interference}
Having said that, there is irreducible interference that we should take into account. Once we turn on the Planck constant, the uncertainty principle creates a width for a path and thus the paths around the original path are summed together. If these paths have different phases, they would cancel out to some extent and yield an additional suppression factor. This is also supported by the fact that the stationary conditions for $\Theta^{(0)}$ reproduce the equations of motion as we have seen in the previous subsection.
Thus, it is preferable to evaluate the tunneling rate at a saddle point of $\Theta^{(0)}$.

Let us clarify what we mean by a saddle point of $\Re\Theta^{(0)}$, which is actually a little obscure unlike that of $\Im\Theta^{(0)}$.
There are two kinds of regions that contribute to $\Re\Theta^{(0)}$; one is the region with $\mathcal K>0$ and $\mathcal V<0$, and the other is that with $\mathcal K<0$ and $\mathcal V>0$. They are separated by a region with $\alpha^2\leq0$. Although their contributions to $\Re\Theta^{(0)}$ have opposite signs with Eq.~\eqref{eq_theta_final}, we need a more careful discussion. 
Let us go back to Eq.~\eqref{eq_alpha_def}.
Although the branch cuts of $\alpha$ are chosen so that it describes a tunneling process for the region with $\alpha^2<0$, there are always two solutions for the region with $\alpha^2>0$. The amplitudes of the two solutions are determined by the junction conditions at the initial state, at the final state, or at the surface of $\mathcal V=0$ or $\mathcal K=0$, but not by Eq.~\eqref{eq_alpha_def}.

If a region with $\alpha^2>0$ is connected to the initial state or the final state, we can choose one of the solutions that represents forward waves and thus we may use the same branch cuts as in Eq.~\eqref{eq_alpha_def}.
However, when a region with $\alpha^2>0$ is surrounded by a region with $\alpha^2\leq0$ and is isolated, the two solutions are summed with certain coefficients and behave like standing waves. The treatment of such a region requires a more rigid definition of tunneling rates beyond the WKB approximation.
Therefore, we restrict the paths so that all the regions with $\alpha^2>0$ are connected to the initial state or to the final state. We also do not consider transitions between states having different signs of $\mathcal K$ to avoid the coexistence of the region with $\mathcal K>0$ and $\mathcal V<0$, and that with $\mathcal K<0$ and $\mathcal V>0$, which has a difficulty in the definition of the relative signs of $\alpha$ between these regions.

In Appendix \ref{apx_re_saddle}, we show that it is actually sufficient to find a minimum of $\Im\Theta^{(0)}$ for the evaluation of a tunneling rate because $|\Re\Theta^{(0)}|$ is always bounded from below. We also demonstrate it numerically in Appendix \ref{apx_osc}.
In addition, even if the minimization of $\Im\Theta^{(0)}$ is incomplete, the result can always be interpreted as a lower bound on the tunneling rate since $\Im\Theta^{(0)}$ is also bounded from below.

\subsubsection{Tunneling Rate}
In summary, we adopt the following definition of the tunneling rate:
\begin{equation}
    \gamma\sim \gamma^{(0)}=\max_{\Phi^M(s,\bfx)}\abs{\frac{\Psi^{(0)}[\Phi^M(s_f,\bfx)]}{\Psi^{(0)}[\Phi^M(s_i,\bfx)]}}^2=\max_{\Phi^M(s,\bfx)}e^{-2\Im\Theta^{(0)}},
\end{equation}
where $\Psi^{(0)}[\Phi^M(s,\bfx)]$ is the leading order WKB wave functional solved from $s_i$ to $s_f$ following path $\Phi^M(s,\bfx)$. The initial configuration is $\Phi^M(s_i,\bfx)$, where $\alpha^2 \geq 0$ is realized everywhere, and the final configuration is $\Phi^M(s_f,\bfx)$, where $\alpha^2$ becomes positive again.
Wherever $\alpha^2<0$ in the path, we take the decaying solution. We do not allow a path that has a region with $\alpha^2>0$ surrounded by a region with $\alpha^2\leq0$. We also do not consider transitions between states having different signs of $\mathcal K$.

Notice that the tunneling rate should have mass-dimension four and we need a dimensionful overall factor. Its evaluation is, however, beyond the scope of this paper. 
Notice also that the WKB approximation would break where the sign of $\alpha^2$ changes, but it only gives $\order{1}$ corrections to the overall factor and thus we ignore such effects.
In addition, we do not take into account the evolution of the bubble after nucleation; since the completion of a phase transition is a more complicated process involving bubble collisions, we focus on a tunneling rate itself.

\subsubsection{Comparison with the CDL Formula}
It is instructive to see the difference between our formula and the widely-used CDL formula. We can reproduce the latter as
\begin{align}
    \Theta^{(0)}&=\int\dd[4]{x}2\sqrt{h}\sqrt{-\mathcal V}\sqrt{\mathcal K}\nonumber\\
    &=\int\dd[4]{x}2\alpha\sqrt{h}(-\mathcal V)\nonumber\\
    &=\int\dd[4]{x}\alpha\sqrt{h}\qty(-\mathcal V+\alpha^{-2}\mathcal K)\nonumber\\
    &=\int\dd[4]{x}\alpha\sqrt{h}\qty[\frac{1}{2\kappa}\qty({}^{(3)}\mathcal R+K_{ij}K^{ij}-K^2)-\frac{g^{\mu\nu}}{2}(\partial_\mu\phi)(\partial_\nu\phi)-V(\phi)]\nonumber\\
    &=\frac{S-\bar S}{2},
\end{align}
where
\begin{align}
    \frac{S}{2}&=\int_{s_i}^{s_f}\dd{s}\int\dd[3]{x}\alpha\sqrt{h}\qty[\frac{1}{2\kappa}\mathcal R-\frac{g^{\mu\nu}}{2}(\partial_\mu\phi)(\partial_\nu\phi)-V(\phi)],\\
    \frac{\bar S}{2}&=-\frac{1}{2\kappa}\int_{s_i}^{s_f}\dd{s}\int\dd[3]{x}\alpha\sqrt{h}D_\rho \mathcal X^\rho. \label{boundary_action}
\end{align}
Here, $\Theta^{(0)}$ gives a half of the CDL action since we stop the $s$-integration at the turning point (the moment when the bubble nucleates).
As we can see clearly, our formula is the same as the CDL one except that it does not depend on $N$ or $N_i$, and
that $\bar S$ automatically determines what we need to subtract from the bounce action. Notice that the action cannot be defined at a point where $\mathcal V=0$ or $\mathcal K=0$, but Eq.~\eqref{eq_theta_final} remains regular and can treat such a point.
A further discussion on the CDL bounce is provided in Section \ref{sec_cdl}.

It is worth noting that our formula is applicable to paths that are not $O(3)$ symmetric, to paths that do not satisfy the stationary conditions, and even to paths that have both Euclidean and Lorentzian evolution. This is distinct from what have been obtained in the previous works and allows us to consider a new possibility of quantum tunneling as we will discuss in Section \ref{sec_wall_tunneling}.

There is also an implication on the CDL formulation; a saddle point of the Euclidean action should be searched among the (four-dimensional) configurations that satisfy the classical Hamiltonian constraint and the classical momentum constraints, but not among all the possible configurations.

\subsubsection{Other Forms of $\bar S$}
Before closing this section, we show other expressions of $\bar S$.
Defining a vector field, $u^{\mu}$, pointing toward the field deformation with $u^{\mu}u_{\mu}=-1$, $\bar S$ can be rewritten as
\begin{equation}
    \frac{\bar S}{2}=-\frac{1}{\kappa}\int_{s_i}^{s_f}\dd{s}\int\dd[3]{x}\alpha\sqrt{h}D_\mu(u^\rho D_\rho u^\mu-u^\mu D_\rho u^\rho),\label{eq_bdy2}
\end{equation}
which recovers the four-dimensional diffeomorphism. The choice of a coordinate system in Eq.~\eqref{boundary_action} corresponds to that of $u^\mu$ in Eq.~\eqref{eq_bdy2}.
 
In Appendix \ref{apx_shift}, we give formulas with the shift functions and find that $\bar S$ is not given by Eq.~\eqref{eq_bdy2}, but by
\begin{equation}
    \frac{\bar S}{2}=-\int_{s_i}^{s_f}\dd{s}\int\dd[3]{x}\partial_\mu\mathcal H_{\rm bdy}^\mu=-\int_{s_i}^{s_f}\dd{s}\int\dd[3]{x}H,
\end{equation}
where $\mathcal H_{\rm bdy}^\mu$ and $H$ are those with $\Phi^M(\bfx)\to\Phi^M(s,\bfx)$, $N(\bfx)\to\alpha(s,\bfx)$, $N_i(\bfx)\to\beta_i(s,\bfx)$, $\pi_N(\bfx)\to0$ and $\pi_N^i(\bfx)\to0$. Here, $\beta_i(s,\bfx)$'s are the functions defined in Appendix \ref{apx_shift}. This expression is supported by the discussions in \cite{Tanaka:1992zw,Gen:1999gi}. Notice that the total derivative terms in $H$ are what we have defined in Eqs.~\eqref{eq_h_bdy1} and \eqref{eq_h_bdy2} without adding any additional terms.

\section{Mini-superspace}\label{sec_minisuperspace}
\setcounter{equation}{0}
We first examine the simplest example with a homogeneous and isotropic configuration.

We make the following Ansatz;
\begin{align}
    h_{ij}&=a^2(s)\bar h_{ij},\\
    \phi&=\phi(s),
\end{align}
where $\bar h_{ij}$ is $s$-independent. We can find that $\alpha$ is also a function of $s$. We have
\begin{equation}
    \mathcal K=\frac12\qty[(\partial_s\phi)^2-\frac{6}{\kappa}\qty(\frac{\partial_sa}{a})^2].
\end{equation}
As we have commented at the end of Subsection \ref{subsec_wkb}, when $\mathcal K=0$, 
$\phi$ and $a$ cannot move keeping $\mathcal K=0$.

The classical Hamiltonian constraint is given by
\begin{equation}
    0=C^{\mathcal H}=a^3\sqrt{\bar h}\qty[\frac{(\partial_s\phi)^2}{2\alpha^2}-\frac{3}{\alpha^2\kappa}\qty(\frac{\partial_sa}{a})^2+V(\phi)-\frac{{}^{(3)}\bar{\mathcal R}}{2\kappa a^2}],\label{eq_mini_hamiltonian}
\end{equation}
where ${}^{(3)}\bar {\mathcal R}$ is the Ricci scalar for $\bar h_{ij}$.
The classical momentum constraints are satisfied trivially.

The stationary conditions for the metric are given by
\begin{equation}
    0=F_{ij}={}^{(3)}\bar{\mathcal R}_{ij}+\bar h_{ij}\qty[\frac{1}{2\alpha}\partial_s\frac{1}{\alpha}\partial_sa^2+\frac{1}{\alpha^2}\qty(\partial_sa)^2-\kappa a^2 V(\phi)],
\end{equation}
where ${}^{(3)}\bar {\mathcal R}_{ij}$ is the Ricci tensor for $\bar h_{ij}$.
Here, we can see that the metric, $\bar h_{ij}$, has to be taken from the beginning so that
\begin{equation}
    {}^{(3)}\bar{\mathcal R}_{ij}\propto \bar h_{ij}.
\end{equation}
Then, there is only one independent equation,
\begin{equation}
    0=\bar h^{ij}F_{ij}={}^{(3)}\mathcal R+3\qty[\frac{1}{2\alpha}\partial_s\frac{1}{\alpha}\partial_sa^2+\frac{1}{\alpha^2}\qty(\partial_sa)^2-\kappa a^2V(\phi)].\label{eq_mini_metric}
\end{equation}
The stationary condition for the scalar field is given by
\begin{equation}
    0=F_\phi=\frac{1}{\alpha a^3}\partial_s\frac{a^3}{\alpha}\partial_s\phi+\dv{V}{\phi}.\label{eq_mini_scalar}
\end{equation}
Notice that Eqs.~\eqref{eq_mini_hamiltonian}, \eqref{eq_mini_metric} and \eqref{eq_mini_scalar} are not independent.

To calculate a tunneling rate, we need a solution of these equations with an initial state and a final state that are outside of the tunnel, {\it i.e.} $\mathcal \alpha^2\geq0$ at $s=s_i$ and at $s=s_f$.
Once such a solution is obtained, the tunneling rate is given by
\begin{equation}
    \ln\gamma^{(0)}=-2\Im\int\dd[3]{x}\sqrt{\bar h}\int_{s_i}^{s_f}\dd{s}\frac{a^3}{\alpha}\qty[(\partial_s\phi)^2-\frac{6}{\kappa}\qty(\frac{\partial_sa}{a})^2].
\end{equation}
We note that, to obtain a finite tunneling rate, the spatial volume has to be finite.

Let us see what has been subtracted from the bounce action. It is
\begin{equation}
   \frac{\bar S}{2}=\frac{3}{\kappa}\int\dd[3]{x}\sqrt{\bar h}\qty[\frac{a^3}{\alpha}\frac{\partial_sa}{a}]_{s=s_i}^{s_f},
   \label{boundary_MSS}
\end{equation}
whose imaginary part is zero if the initial state and the final state are outside of the tunnel, {\it i.e.} $\alpha^2\geq0$. Thus, we do not need to subtract anything from the bounce action to obtain the suppression factor of the tunneling rate in this example.
\section{CDL bounce}\label{sec_cdl}
\setcounter{equation}{0}
\subsection{Review on the Coleman-De Luccia bounce and the Hawking-Moss bounce}
\label{sec_review_cdl_hm}
We begin by briefly reviewing the tunneling rate by Coleman and De Luccia \cite{Coleman:1980aw} and that by Hawking and Moss \cite{Hawking:1981fz}.

In the CDL formulation, the tunneling rate is calculated from the CDL bounce, which is an $O(4)$ symmetric solution to the Euclidean equations of motion, Eqs.~\eqref{eq_einstein} and \eqref{eq_scalar_eom}. With radial coordinate $\xi$, we adopt the following $O(4)$-symmetric Ansatz:
\begin{align}
    \phi&=\phi(\xi),\\
    g_{\mu\nu}\dd{x^\mu} \dd{x^\nu}&=\dd{\xi}^2+\rho^2(\xi)[\dd{\chi}^2+\sin^2\chi(\dd{\theta}^2+\sin^2\theta \dd{\phi}^2)].\label{eq_o4_metric}
\end{align}
Then, the equations of motion for $\phi$ and $\rho$ reduce to
\begin{align}
     & \phi''+\frac{3\rho'}{\rho}\phi'=\pdv{V}{\phi},\label{scalar_eom}               \\
     & \rho'^2=1+\frac{\kappa}{3}\rho^2\qty(\frac{1}{2}\phi'^2-V(\phi)),
     \label{rho_eom}
\end{align}
where the prime indicates the derivative with respect to $\xi$.
The bounce should also satisfy the following boundary conditions:
\begin{align}
    \phi'(0)=0,~\phi(\infty)=\phi_{\rm F},~\rho(0)=0,
\end{align}
or
\begin{align}
    \phi'(0)=0,~\phi'(\xi_{\max})=0,~\rho(0)=0,~\rho(\xi_{\max})=0,
\end{align}
where $\xi=\xi_{\max}$ is the point where $\rho(\xi)$ becomes zero again. The latter conditions are for the de Sitter background.

Once the bounce is obtained, its Euclidean action can be calculated as
\begin{equation}
    S_{\rm E}^{\rm CDL}=2\pi^2\int d\xi\qty[\rho^3\qty(\frac{1}{2}\phi'^2+V(\phi))+\frac{3}{\kappa}(\rho^2\rho''+\rho\rho'^2-\rho)],\label{eq_cdl_action}
\end{equation}
and the tunneling rate is given by
\begin{equation}
    \gamma=\mathcal Ae^{-(S_{\rm E}^{\rm CDL}-S_{\rm E}^{\rm FV})},\label{eq_cdl_rate}
\end{equation}
where $\mathcal A$ is a pre-factor and $S_{\rm E}^{\rm FV}$ is the Euclidean action at the false vacuum.

The Hawking-Moss (HM) bounce is analogous to the CDL bounce but with a spatially homogeneous configuration at the top, $\phi_{\rm top}$, of the potential barrier. Then, the exponent of Eq.~\eqref{eq_cdl_rate} is replaced by
\begin{align}
    S_{\rm E}^{\rm HM}-S_{\rm E}^{\rm FV}&=-\frac{24\pi^2}{\kappa^2V(\phi_{\rm top})}+\frac{24\pi^2}{\kappa^2V(\phi_{\rm F})}\nonumber\\
    &\simeq \frac{\Delta M}{T_{\rm F}},\label{eq_hm_action}
\end{align}
where the last line gives an approximation for $\Delta V= V(\phi_{\rm top})-V(\phi_{\rm F})\ll V(\phi_{\rm F})$ and we have assumed $V(\phi_{\rm F})>0$. Here, $\Delta M = (4 \pi/3) (1/H_{\rm F})^3 \Delta V$ with $H_{\rm F} = \sqrt{\kappa V(\phi_{\rm F})/3}$, and $T_{\rm F} = H_{\rm F}/(2 \pi)$ is the Hawking temperature of the background de Sitter space. From the last expression, we can see that it can be understood as the thermal jump from the false vacuum to the top of the potential. In the following, we use ``thermal jump'' to describe the suppression factor of the form of Eq.~\eqref{eq_hm_action}, but there is no fundamental separation between quantum and thermal processes.

\begin{figure}[t]
  \begin{center}
    \includegraphics[width=0.4\linewidth]{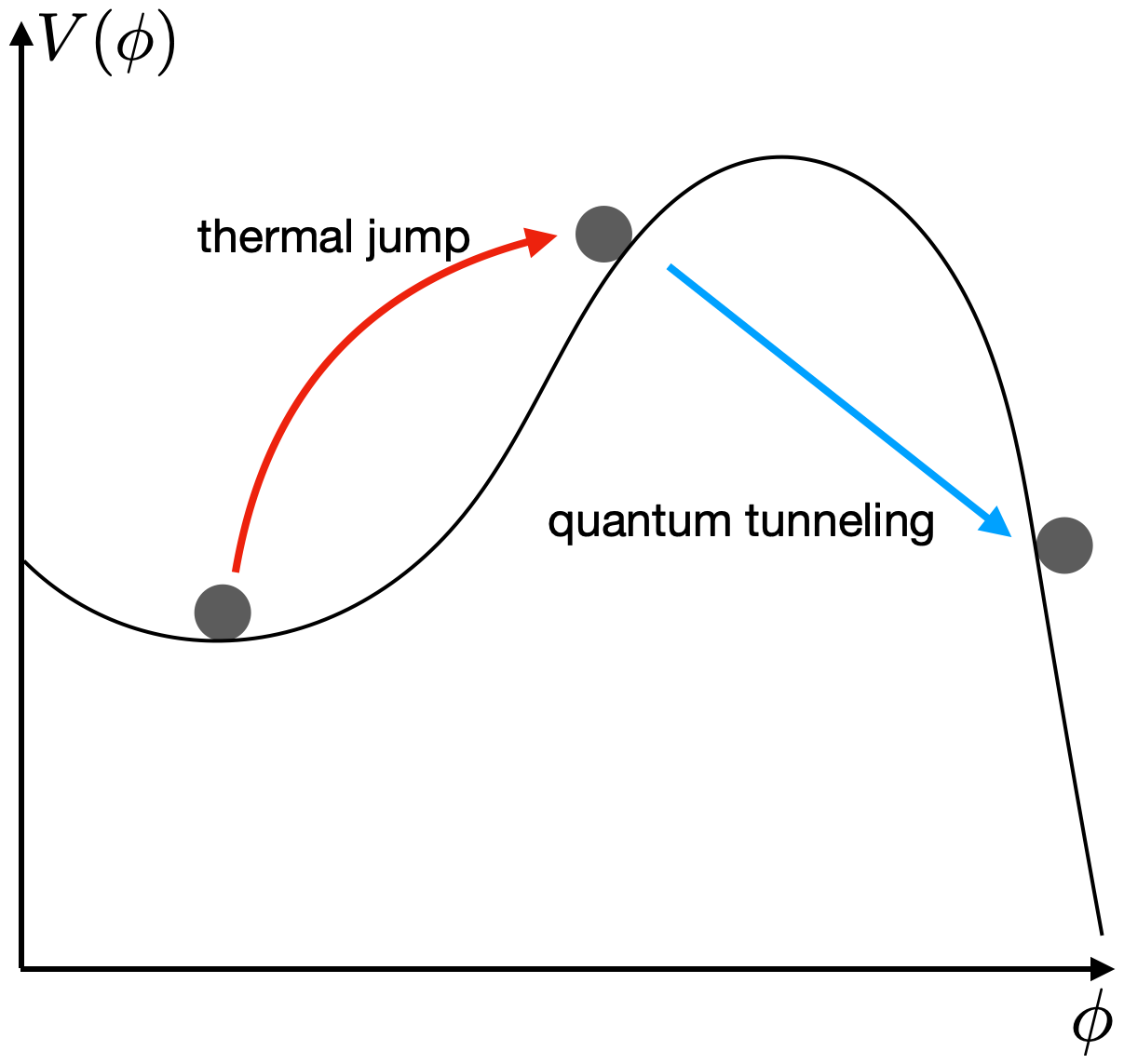}
  \end{center}
\caption{Schematic picture of a phase transition that occurs in a combination of a thermal jump and quantum tunneling.
}
\label{thermal_quantum_pic}
\end{figure}

Though the HM bounce always exists for the decay of a de Sitter vacuum, it is not the case for the CDL bounce. A useful quantity to check for its existence is
\begin{equation}
\tilde \beta = -\frac{3}{\kappa V(\phi)} \left. \frac{d^2 V}{d\phi^2} \right|_{\phi = \phi_{\rm top}}.\label{eq_beta_tilde}
\end{equation}
If $\tilde \beta$ is much larger than 4, there exists a CDL bounce and it describes the transition from a configuration close to the false vacuum. As $\tilde\beta$ gets closer to 4, $\phi (0)$ and $\phi (\xi_{\max})$ get closer to each other. In this regime, the CDL tunneling formula describes the transition occurring in a combination of the thermal jump from $\phi_{\rm F}$ to $\phi (\xi_{\max})$ and the quantum tunneling from $\phi (\xi_{\max})$ to the turning point (see Fig.~\ref{thermal_quantum_pic}).
Then, at $\tilde\beta=4$, the CDL bounce localizes at the top of the potential barrier and there is no CDL bounce for $\tilde\beta < 4$. 

In the following subsection, we try to use the CDL bounce in our formula. It means that the initial state is taken to be $\phi (\xi_{\max})$, but not $\phi_F$. Thus, we expect that our formula gives only the transition rate for the quantum tunneling part, {\it i.e.}
\begin{equation}
    \gamma=\mathcal Ae^{-(S_{\rm E}^{\rm CDL}-S_{\rm E}^{\rm hor})},\label{eq_our_rate}
\end{equation}
where $S_{\rm E}^{\rm hor}$ is the minus Bekenstein-Hawking entropy of the cosmological horizon,
\begin{align}
    S_{\rm E}^{\rm hor}=-\frac{24\pi^2}{\kappa^2V(\phi(\xi_{\max}))}.
\end{align}

\begin{figure}[t]
  \begin{center}
    \includegraphics[keepaspectratio=true,height=53mm]{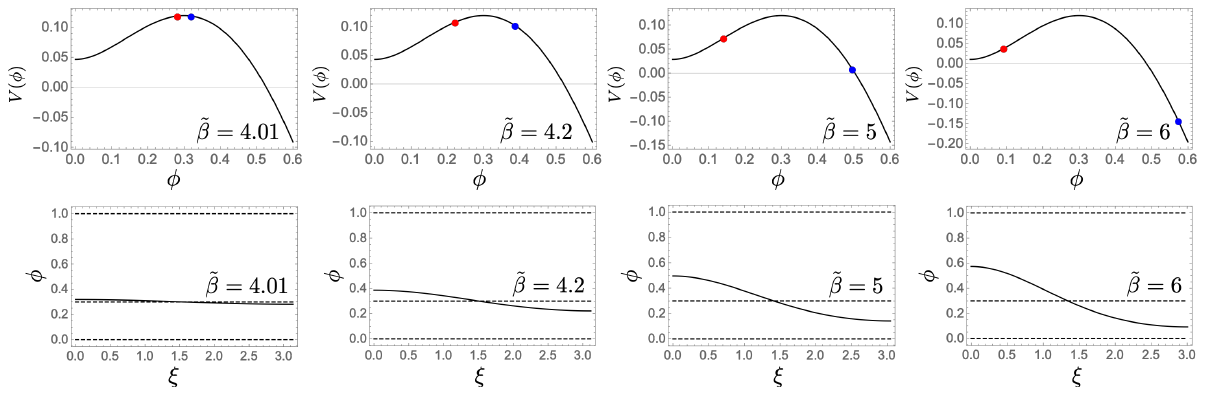}
  \end{center}
\caption{
The shape of the potential and the CDL bounce. The top panels show the normalized potential, where the red points and the blue points indicate $\phi= \phi(0)$ and $\phi(\xi_{\rm max})$, respectively. In the bottom panels, the solid line shows the CDL bounce and the dashed lines indicate the positions of the top and the bottom of the potential.
}
\label{cdl_bounce_pic}
\end{figure}

\begin{figure}[t]
  \begin{center}
    \includegraphics[width=0.9\linewidth]{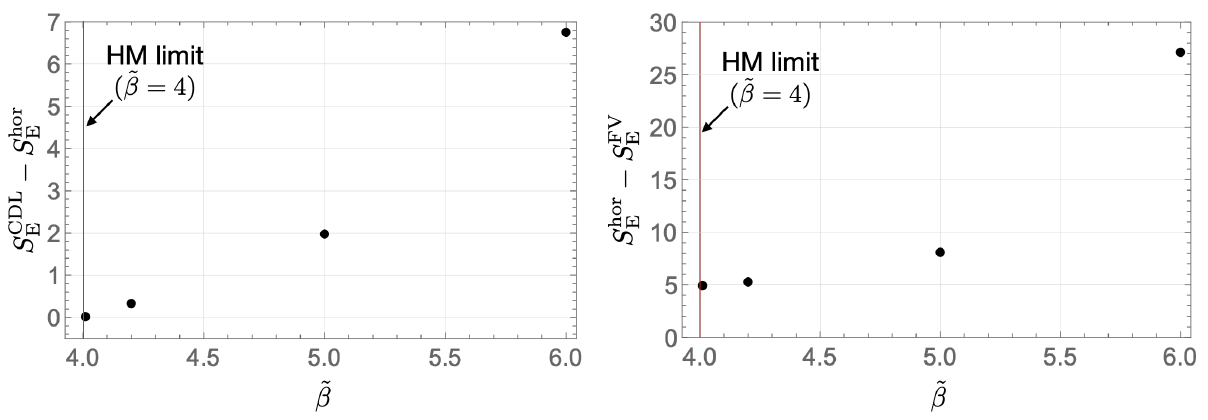}
  \end{center}
\caption{
Plot of $S_{\rm E}^{\rm CDL}- S_{\rm E}^{\rm hor}$ (left) and $S_{\rm E}^{\rm hor}- S_{\rm E}^{\rm FV}$ (right) with respect to $\tilde{\beta}$.
}
\label{action_beta_pic}
\end{figure}

Lastly, we demonstrate the computation of the tunneling rate. We adopt the following potential for the scalar field:
\begin{equation}
    V(\phi)=\frac{\tilde\beta}{k(1-k)}\qty(\frac{1}{4}\phi^4-\frac{k+1}{3}\phi^3+\frac{k}{2}\phi^2-\frac{k^3}{12}(2-k))+\frac{3}{\kappa},
\end{equation}
where $0<k<0.5$ is a constant and $\tilde\beta$ parametrizes the curvature of the potential at the top, which is the same parameter as in Eq.~\eqref{eq_beta_tilde}. We set the Planck mass to one, {\it i.e. $\kappa=8\pi$}. The bounce is solved numerically using the under-shoot over-shoot method and the solutions are shown in Fig.~\ref{cdl_bounce_pic}. The parameters and the values of the action are given in Table \ref{table_action}.
In Fig.~\ref{action_beta_pic}, we plot the values of $S_{\rm E}^{\rm CDL}-S_{\rm E}^{\rm hor}$ and $S_{\rm E}^{\rm hor}-S_{\rm E}^{\rm FV}$ with respect to $\tilde\beta$. The former is the suppression factor for the quantum tunneling and the latter is that for the thermal jump. As $\tilde\beta\to4$, the quantum tunneling rate approaches zero and the transition rate is determined by the rate of the thermal jump.

\begin{table}[t]
\begin{center}
\begin{tabular}{c c c c c c} 
\hline
 $\tilde\beta$ & $S_{\rm E}^{\rm CDL}$ & $S_{\rm E}^{\rm hor}$ & $S_{\rm E}^{\rm FV}$ & $S_{\rm E}^{\rm CDL}-S_{\rm E}^{\rm hor}$ & $S_{\rm E}^{\rm hor}-S_{\rm E}^{\rm FV}$\\ [0.5ex] 
\hline\hline
6 & $-3.3114$ & $-10.0651$ & $-37.2006$ & $6.7537$ & $27.1355$\\
 \hline
5 & $-3.1847$ & $-5.1600$ & $-13.2533$ & $1.9753$ & $8.0933$ \\
 \hline
4.2 & $-3.1434$ & $-3.4740$ & $-8.7482$ & $0.33062$ & $5.2741$\\
 \hline
4.01 & $-3.1416$ & $-3.1592$ & $ -8.0947$ & $0.017596$ & $4.9355$
\end{tabular}
\end{center}
\caption{The parameters and the values of action. We take $\kappa=8\pi$ and $k=0.3$.}
\label{table_action}
\end{table}

\subsection{CDL bounce in our formulation}
Let us calculate the decay rate using our formulation and the CDL bounce. It is crucial to choose the $s$-direction so that it parametrizes the field deformation, and to set the initial state and the final state so that they are outside of the tunnel, {\it i.e.} $\alpha^2\geq0$ everywhere. The CDL bounce can satisfy these conditions in the static coordinates, where the imaginary time parametrizes the field evolution between them and the time derivatives of $\phi$ and $h_{ij}$ vanish at the initial state and the final state (the turning point).

Let us first execute a coordinate transformation from the $O(4)$ global coordinates, $(\xi,\chi,\theta,\phi)$, to the $O(3)$ static coordinates, $(s,r,\theta,\phi)$. 
We parametrize the metric in the static coordinates as
\begin{equation}
    g_{\mu\nu}\dd{x^\mu} \dd{x^\nu}=-\alpha^2(s,r)\dd{s}^2+A^2(s,r)\dd{r}^2+r^2(\dd{\theta}^2+\sin^2\theta \dd{\phi}^2).
    \label{static_coordinates}
\end{equation}

Since we use the same $\theta$ and $\phi$ in both coordinate systems, we have
\begin{equation}
    r=\rho(\xi)\sin\chi.
\end{equation}

To make $g_{s r}=0$, we need
\begin{equation}
    \pdv{\xi}{s}\pdv{\xi}{r}+\rho^2\pdv{\chi}{s}\pdv{\chi}{r}=0,
\end{equation}
which is equivalent to
\begin{equation}
    \pdv{s}{\chi}=-\rho\rho'\tan\chi\pdv{s}{\xi}.
\end{equation}
Its general solution is given by
\begin{equation}
    s=T\qty(f(\xi)\cos\chi),
\end{equation}
where $T(x)$ is an arbitrary smooth injective function and $f(\xi)$ is the solution of
\begin{equation}
    f'(\xi)=\frac{f(\xi)}{\rho(\xi)\rho'(\xi)},
\end{equation}
with
\begin{equation}
    \lim_{\xi\to0}\frac{f(\xi)}{\xi}=1.
\end{equation}
We take $T(x)$ so that $(s,r)=(s_f,0)$ corresponds to $\xi=0$, and $(s,r)=(s_i,0)$ corresponds to $\xi=\infty$ or $\xi_{\max}$.
Then, applying the transformation to Eq.~\eqref{eq_o4_metric}, we obtain\footnote{
Notice that Eq.~\eqref{eq_alpha_def} holds independently of the choice of $T(x)$ for the solution of the equations of motion.}
\begin{align}
    A^2&=\qty(1+\frac{\rho'^2-1}{\rho^2}r^2)^{-1},\\
    \alpha^2&=-\qty(1+\frac{\rho'^2-1}{\rho^2}r^2)^{-1}\qty(\frac{\rho\rho'}{T'f})^2.\label{eq_alpha_cdl}
\end{align}

Let us evaluate the tunneling rate.
One can easily show
\begin{align}
    G^{ijkl}(\partial_sh_{ij})(\partial_sh_{kl})&=0,\\
    g^{ss}(\partial_s\phi)^2&=-\frac{(\phi')^2}{1+\rho'^2\tan^2\chi}.
\end{align}
From these equations and the Hamiltonian constraint, we see $\mathcal V$ and $\mathcal K$ are positive definite and thus $\Im\alpha$ is negative definite. Then, from Eq.~\eqref{theta_var_2}, we get
\begin{align}
    \ln\gamma^{(0)}&=-2\int^{s_f}_{s_i}\dd{s}\int\dd[3]{x}|\alpha|\sqrt{h}\frac{(\phi')^2}{1+\rho'^2\tan^2\chi}\nonumber\\
    &=-8\pi\int \dd{\xi}\int^{\pi/2}_0 \dd{\chi} \rho^3\sin^2\chi\frac{(\phi')^2}{1+\rho'^2\tan^2\chi}\nonumber\\
    &=-2\pi^2\int \dd{\xi}\rho^3\qty(\frac{\phi'}{1+|\rho'|})^2.
    \label{wdw_decay_exponent}
\end{align}

Let us see if it agrees with Eq.~\eqref{eq_our_rate}.
The vector, $u^\mu$, in Eq.~\eqref{eq_bdy2} in the static coordinates is given by
\begin{equation}
    u^\mu=\qty(\frac{i}{|\alpha|},0,0,0).
\end{equation}
It can be transformed into the global coordinates as
\begin{equation}
    u^\mu=\qty(\frac{i}{\sqrt{1+\rho'^2\tan^2\chi}},-\frac{\rho'\tan\chi}{\rho}\frac{i}{\sqrt{1+\rho'^2\tan^2\chi}},0,0).
\end{equation}
We get
\begin{align}
    \bar S_{\rm E}&=-i\bar S\nonumber\\
    &=\frac{2}{\kappa}\int_{s_i}^{s_f}\dd{s}\int\dd[3]{x}|\alpha|\sqrt{h}D_\mu(u^\rho D_\rho u^\mu-u^\mu D_\rho u^\rho)\nonumber\\
    &=\frac{8\pi}{\kappa}\int\dd{\xi}\int^{\pi/2}_0\dd{\chi}\rho^3\sin^2\chi D_\mu(u^\rho D_\rho u^\mu-u^\mu D_\rho u^\rho)\nonumber\\
    &=\eval{-\frac{2\pi^2 }{\kappa}\frac{\rho^2}{1+\rho'}(2-\rho'-\rho'^2)}_{\xi=\xi_{\max}}\nonumber\\
    &=S_{\rm E}^{\rm hor}.
\end{align}
Here, we have used $\rho'\sim-1+\frac{\kappa}{6}V(\phi)\rho^2$ at $\xi\sim \xi_{\max}$. Thus, the tunneling rate can be expressed as in Eq.~\eqref{eq_our_rate}; if we use the CDL bounce in our formula, it gives only the quantum tunneling part. This is because the CDL bounce with the hyperspherical coordinates sets the initial state as $\phi(s_i,\bfx)=\phi(\xi_{\max})$. The evaluation of the full tunneling rate using our formulation would require a path that has an initial state, $\phi(s_i,\bfx)=\phi_{\rm F}$.

Notice that $\Im\bar S$ is generally non-zero in this example, which is in contrast with the mini-superspace example. 
These two examples demonstrate that the tunneling rate indeed depends on the direction of the field deformation, which is determined by the choice of the coordinate system in Eq.~\eqref{boundary_action}, or equivalently by that of $u^\mu$ in Eq.~\eqref{eq_bdy2}. 

\section{Polychronic Tunneling}\label{sec_wall_tunneling}
\setcounter{equation}{0}
The most distinctive feature of our formulation is that $\alpha^2$ can change its sign locally unlike in the CDL formulation. It opens up a possibility of the tunneling process that involves both Euclidean and Lorentzian evolution, which cannot be considered in the CDL formulation since $\alpha^2$ is negative definite as we can see from Eq.~\eqref{eq_alpha_cdl}.

We consider a path with two degrees of freedom\footnote{These are sufficient to solve all the equations of motion. We have a classical Hamiltonian constraint, a classical momentum constraint, and three stationary conditions. Two of them are not independent. We use the classical Hamiltonian constraint to solve $\alpha$ and the classical momentum constraint to solve $\eta$. The remaining one gives the equations of motion for $\phi$.}, $\phi(s,r)$ and $\eta(s,r)$;
\begin{align}
    \phi&=\phi(s,r),\\
    h_{ij}\dd{x^i} \dd{x^j}&=e^{\eta(s,r)}\dd{r}^2+r^2(\dd{\theta}^2+\sin^2\theta \dd{\phi}^2),
\end{align}
which give
\begin{align}
    \mathcal K&=\frac{1}{2}(\partial_s\phi)^2,\label{eq_cal_k_wt}\\
    \mathcal V&=V(\phi)+\frac{e^{-\eta}}{2}(\partial_r\phi)^2-\frac{1}{\kappa}\qty(\frac{1-e^{-\eta}}{r^2}+\frac{e^{-\eta}}{r}\partial_r\eta).\label{eq_cal_v_wt}
\end{align}
Here, we adopt
\begin{equation}
    V(\phi)=\frac{\phi^4}{4}-\frac{k+1}{3}\phi^3+\frac{k}{2}\phi^2+V_0,\label{eq_potential}
\end{equation}
where $0<k<0.5$ and $V_0$ are constants. The units for the dimensionful quantities are determined by the value of $\kappa$. Notice that $\mathcal K$ is positive definite and thus we do not need to worry about the subtlety at $\mathcal K=0$ discussed at the end of Subsection \ref{subsec_wkb}.

For the initial conditions, we take
\begin{align}
    \phi(0,r)&=\phi_{\rm F},\\
    \eta(0,r)&=-\ln\qty(1-\frac{\kappa r^2}{3}V(\phi_{\rm F})).
\end{align}
Here, we took $\eta(0,r)$ so that the initial state is at the endpoint of the tunnel, {\it i.e.} $\mathcal V=0$. Notice that, from the classical Hamiltonian constraint, it also means $(\pi^{(0)}_\phi)^2=2h\mathcal K/\alpha^2=0$, where $\pi^{(0)}_\phi$ is the classical momentum for $\phi$.

To avoid the numerical difficulties around the pole of $h_{rr}$, we concentrate on the case where the field deformation occurs within $r<r_{\rm max}$ with $r_{\rm max}\ll \sqrt{3/\kappa/V(\phi_F)}$. In this regime, the contribution of the thermal jump process is negligible.

As we have explained in Section \ref{sec_formulation}, a WKB solution can be constructed with an arbitrary path that satisfies the momentum constraints; there is only one non-trivial constraint,
\begin{equation}
    \partial_s\eta=\kappa r(\partial_r\phi)(\partial_s\phi).\label{eq_eta_evol}
\end{equation}
Thus, for given $\phi(s,r)$ and $\eta(0,r)$, we can determine $\eta(s,r)$ uniquely. Since a change of $\phi(s,r)$ in $0<s<1$ affects $\eta(s_f,r)$, we do not fix the final state.

We consider a set of $\phi(s,r)$'s having non-trivial final states that satisfy $\alpha^2>0$ everywhere. Within this set, we maximize the following functional\footnote{As we explain in Appendices \ref{apx_re_saddle} and \ref{apx_osc}, it is enough to optimize only $\Im\Theta^{(0)}$.}:
\begin{align}
    \ln\gamma^{(0)}[\phi(s,r)]&=-16\pi\int_{0}^{1}\dd{s}\int_0^{r_{\max}}\dd{r}I(s,r),
\end{align}
where
\begin{align}
    I(s,r)&=r^2e^{\eta/2}\sqrt{\mathcal K}\Re\qty(\sqrt{\mathcal V}).
\end{align}
Here, we rescaled $s$ so that $s=0$ corresponds to the initial state and $s=1$ to the final state.

We execute Monte Carlo optimization to maximize $\ln\gamma^{(0)}[\phi(s,r)]$ on a $200\times200$ lattice\footnote{We have checked that increasing it to a $300\times300$ lattice does not change our results significantly.} of $(s,r)$. The details of the numerical analysis are given in Appendix \ref{apx_numerical}.

\begin{figure}[t]
  \begin{minipage}{0.49\linewidth}
    \includegraphics[width=\linewidth]{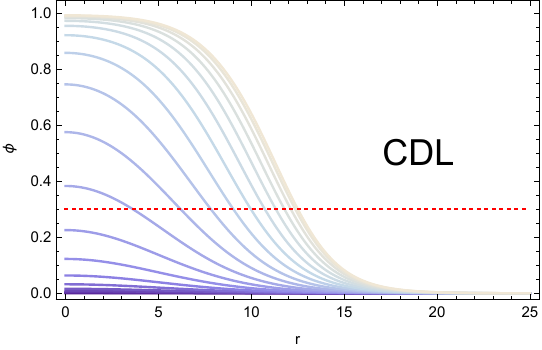}
  \end{minipage}
  \begin{minipage}{0.49\linewidth}
    \includegraphics[width=\linewidth]{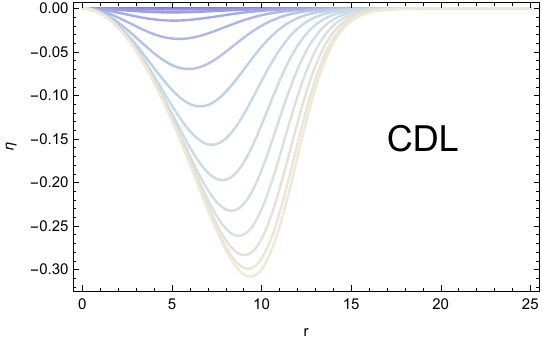}
  \end{minipage}
  \begin{minipage}{0.49\linewidth}
    \includegraphics[width=\linewidth]{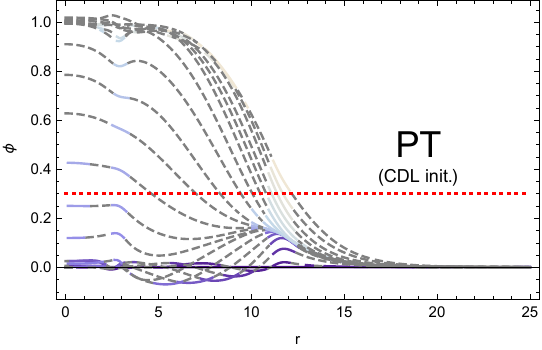}
  \end{minipage}
  \begin{minipage}{0.49\linewidth}
    \includegraphics[width=\linewidth]{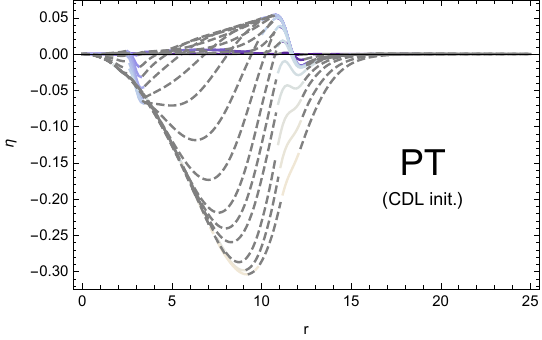}
  \end{minipage}
  \begin{minipage}{0.49\linewidth}
    \includegraphics[width=\linewidth]{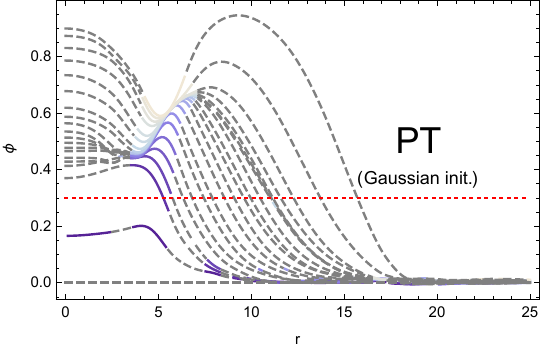}
  \end{minipage}
  \begin{minipage}{0.49\linewidth}
    \includegraphics[width=\linewidth]{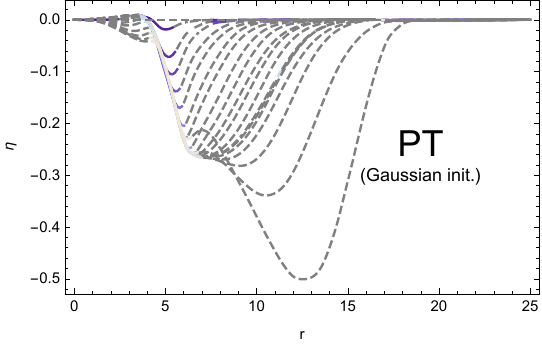}
  \end{minipage}
\caption{The CDL bounce and the PT path with $(\kappa,k,V_0)=(0.5,0.3,0)$. The top panels are for the CDL bounce and the other ones are for the PT path. We take the CDL initial path in the middle panels and the Gaussian initial path in the bottom panels. The left panels are for $\phi$ and the right ones are for $\eta$. Darker blue colors correspond to smaller values of $s$. Where the lines become gray and dashed, $\alpha^2$ is positive. The red dotted line indicates the location of the top of the potential barrier.}
\label{fig_profile_1}
\end{figure}

For the first example, we take $\kappa=0.5$, $k=0.3$ and $V_0=0$. The CDL bounce is shown in the top panels of Fig.~\ref{fig_profile_1}, where the left panel is for $\phi$ and the right one is for $\eta$. The color gradient shows different $s$ and darker blue colors correspond to smaller values of $s$. The interval of the lines is $\Delta s=0.05$, which corresponds to $10$ times the lattice spacing. The CDL bounce gives a tunneling rate of $\ln\gamma^{(0)}\simeq-710$.

\begin{figure}[t]
  \begin{minipage}{0.49\linewidth}
    \includegraphics[width=\linewidth]{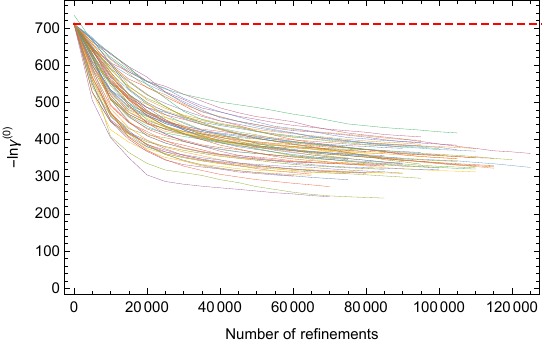}
  \end{minipage}
  \begin{minipage}{0.49\linewidth}
    \includegraphics[width=\linewidth]{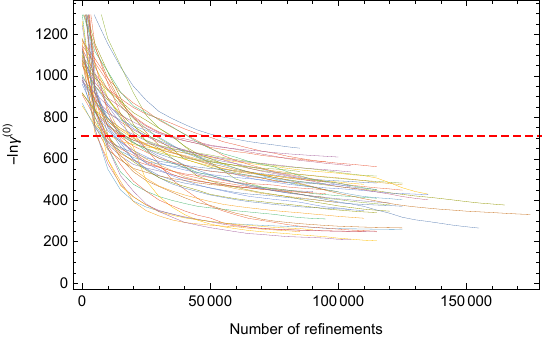}
  \end{minipage}
\caption{Optimization histories with $(\kappa,k,V_0)=(0.5,0.3,0)$. The left panel is with the CDL initial path and right panel is with the Gaussian initial path. The red dashed line indicates the CDL tunneling rate.}
\label{fig_opt_histry}
\end{figure}

We optimize the path taking the CDL bounce as the initial path. The optimization history is shown in the left panel of Fig.~\ref{fig_opt_histry}. The red dashed line shows the tunneling rate of the CDL bounce. Each thin line shows an attempt of the optimization. There are 64 attempts and each one is optimized with a CPU core for about two days.
We can see that there exist paths that have higher tunneling rates than that with the CDL bounce.
The path with the highest tunneling rate is shown in the middle panels of Fig.~\ref{fig_profile_1}. Where the lines become dashed, $\alpha^2$ is positive and hence it corresponds to the region with the Lorentzian evolution.
We call the path the polychronic tunneling (PT) path and it gives a tunneling rate of $\ln\gamma^{(0)}\simeq-242$.

\begin{figure}[t]
\begin{center}
    \includegraphics[width=0.49\linewidth]{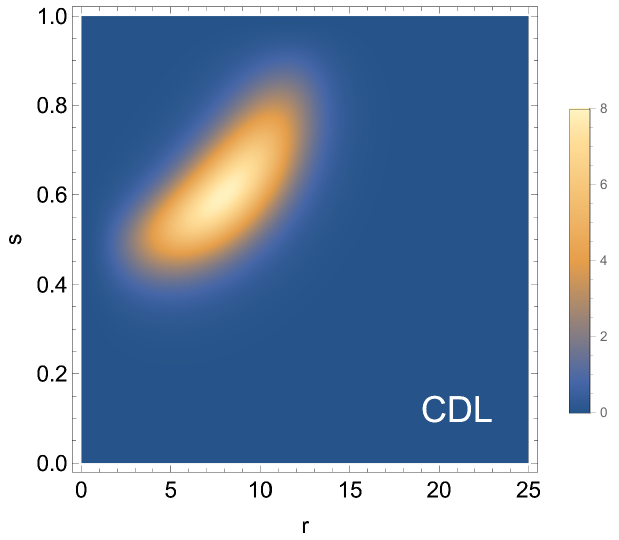}
\end{center}
  \begin{minipage}{0.49\linewidth}
    \includegraphics[width=\linewidth]{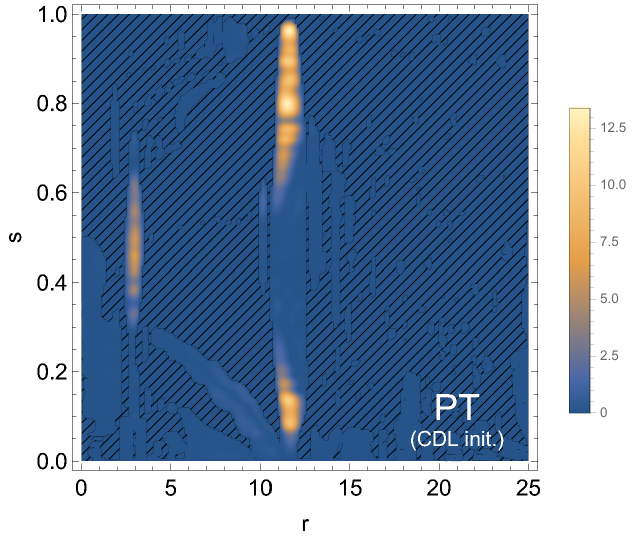}
  \end{minipage}
  \begin{minipage}{0.49\linewidth}
    \includegraphics[width=\linewidth]{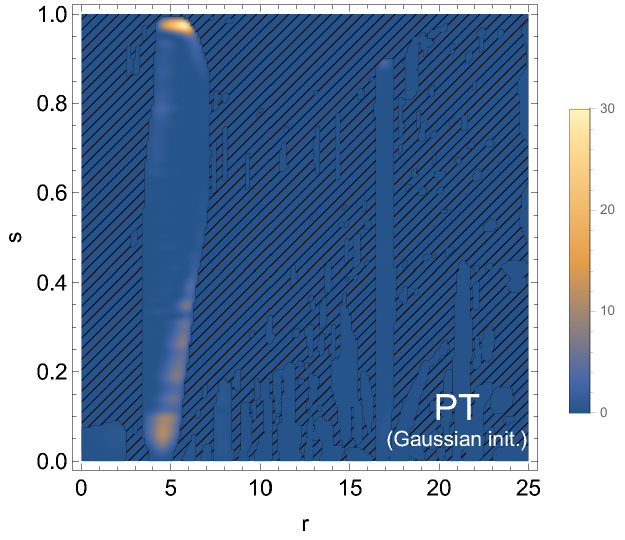}
  \end{minipage}
\caption{The integrand, $I(s,r)$, for the CDL bounce (top) and for the PT path (bottom) with $(\kappa,k,V_0)=(0.5,0.3,0)$. The bottom left panel is with the CDL initial path and the bottom right one is with the Gaussian initial path. The hatched region corresponds to $\alpha^2>0$.}
\label{fig_integrand_1}
\end{figure}

In Fig.~\ref{fig_integrand_1}, we show the integrand, $I(s,r)$, for the CDL bounce (top) and the PT path (bottom left). For the CDL bounce, a small bubble is created and then its wall moves outward until the bubble materializes.
Thus, we have a brighter region diagonally.
On the other hand, for the PT path, the wall around $r\sim12$ mainly tunnels (see also the middle left panel of Fig.~\ref{fig_profile_1}). Around $r\sim3$, we also see a small wall assisting the tunneling, which often appears when we optimize the path starting from the CDL initial path.
The horizontal stripe patterns are due to the freedom to rescale $s$ independently of $r$.
We show $I(s,r)$ for other paths with high tunneling rates in Appendix \ref{apx_others}.

Instead of using the CDL bounce as an initial path, we also try the Gaussian initial paths,
\begin{equation}
    \phi(s,r)=se^{-\frac{r^2}{2\sigma^2}},
\end{equation}
where we take $\sigma\in[5,10]$ randomly. The optimization history is shown in the right panel of Fig.~\ref{fig_opt_histry}. The path with the highest tunneling rate is shown in the bottom panels of Fig.~\ref{fig_profile_1} and its $I(s,r)$ is shown in the bottom right panel of Fig.~\ref{fig_integrand_1}. Its tunneling rate, $\ln\gamma^{(0)}\simeq-204$, is similar to that with the CDL initial path.
Although the deformation of $\phi$ around $s\sim0-0.1$ (the first two gaps) is fast, the actual lattice points interpolate the gaps at regular intervals and there is no numerical issue here.
The dip appearing around $(s,r)\sim(1,5)$ can be understood as follows. After $\phi$ goes over the top of the potential barrier ($\phi=0.3$), the regions with the Lorentzian evolution roll down the potential and eventually take over the tunneling region. Then, the tunneling region is finally pulled by the rolling regions and escapes from the tunnel, which appears as a dip. Notice that since we do not optimize $\Re\Theta^{(0)}$, the path actually does not correspond to a classical path. In Appendix \ref{apx_osc}, we optimize both $\Im\Theta^{(0)}$ and $\Re\Theta^{(0)}$ and observe the similar behavior. 

\begin{figure}[t]
  \begin{minipage}{0.49\linewidth}
    \includegraphics[width=\linewidth]{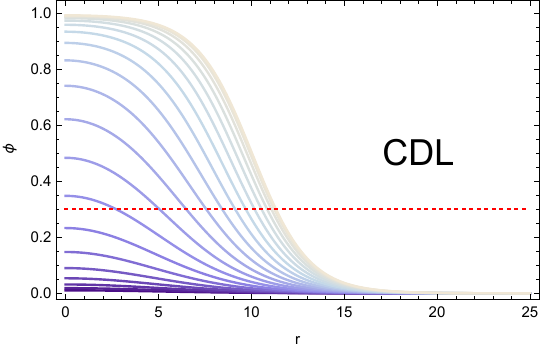}
  \end{minipage}
  \begin{minipage}{0.49\linewidth}
    \includegraphics[width=\linewidth]{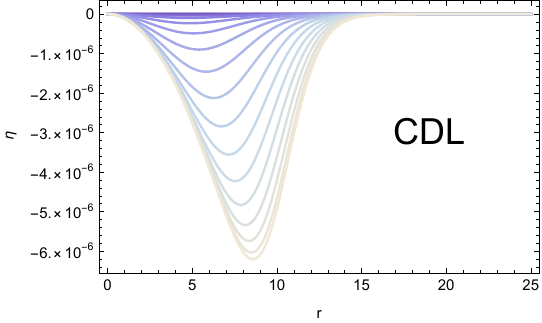}
  \end{minipage}
  \begin{minipage}{0.49\linewidth}
    \includegraphics[width=\linewidth]{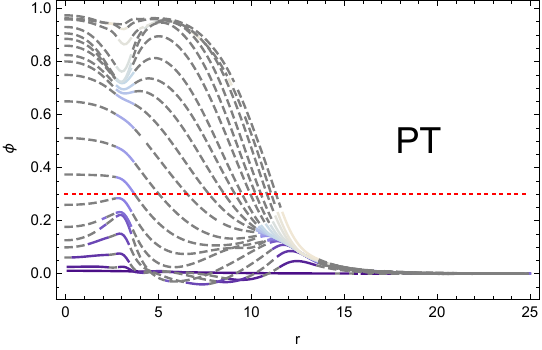}
  \end{minipage}
  \begin{minipage}{0.49\linewidth}
    \includegraphics[width=\linewidth]{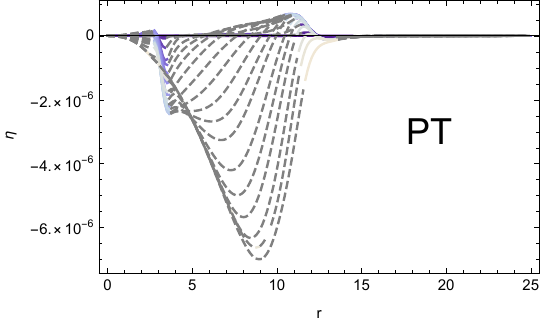}
  \end{minipage}
\caption{The same figure as Fig.~\ref{fig_profile_1} but with $(\kappa,k,V_0)=(10^{-5},0.3,0)$. The top panels are for the CDL bounce and the bottom panels are for the PT path with the CDL initial path.}
\label{fig_profile_2}
\end{figure}

\begin{figure}[t]
  \begin{minipage}{0.49\linewidth}
    \includegraphics[width=\linewidth]{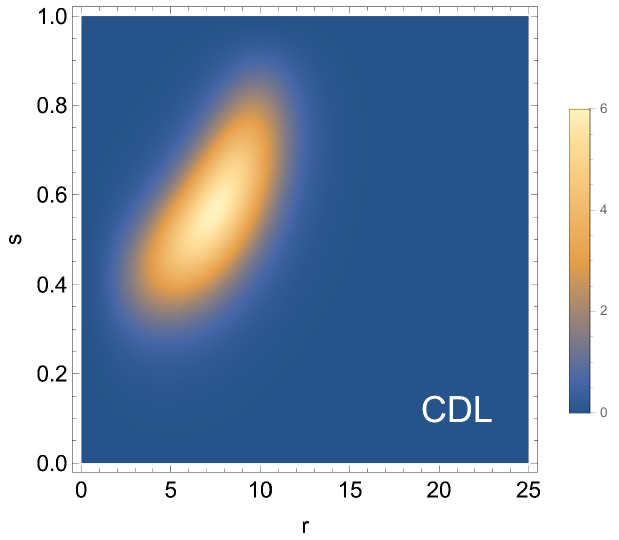}
  \end{minipage}
  \begin{minipage}{0.49\linewidth}
    \includegraphics[width=\linewidth]{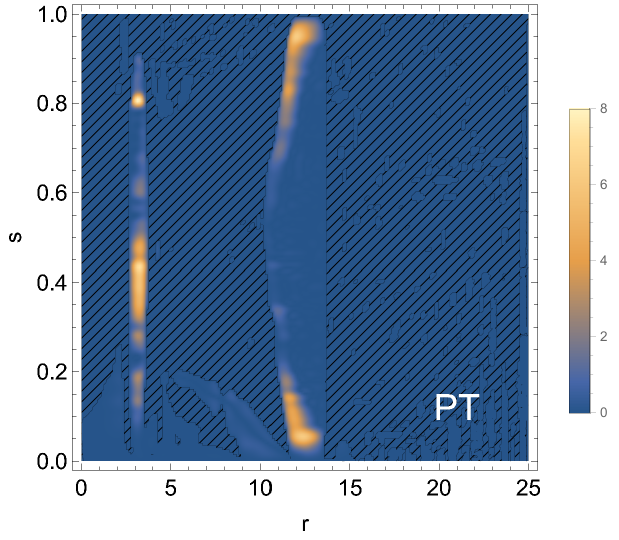}
  \end{minipage}
\caption{The same figure as Fig.~\ref{fig_integrand_1} but with $(\kappa,k,V_0)=(10^{-5},0.3,0)$. The left panel is for the CDL bounce and the right panel is for the PT path with the CDL initial path.}
\label{fig_integrand_2}
\end{figure}

The other example is with $\kappa=10^{-5}$, $k=0.3$ and $V_0=0$, where gravitational effects effectively decouple\footnote{We have checked that taking $\kappa=10^{-8}$ does not change our results significantly.}. We adopt the CDL initial path. The path with the highest tunneling rate and its $I(s,r)$ are shown in Figs.~\ref{fig_profile_2} and \ref{fig_integrand_2} together with those for the CDL bounce. The CDL bounce gives $\ln\gamma^{(0)}\simeq-596$ whereas the PT path gives $\ln\gamma^{(0)}\simeq-154$. Thus, even in the decoupling regime of gravity, the tunneling rate is enhanced very much. This is because the sign of $\alpha^2$ can flip regardless of the size of $\kappa$.
As we can see from Eq.~\eqref{eq_cal_v_wt}, even when the change of $\eta$ is $\order{\kappa}$, its effect on $\mathcal V$ remains $\order{1}$.
In addition, we can imagine that the same thing happens when we connect the CDL solution to the Lorentzian bubble expansion solution, where the sign of $\mathcal V$ flips smoothly at the junction even in the decoupling limit of gravity. We would say that we have just been unaware of the PT paths for decades since the local flip of the sign cannot happen in the CDL formulation.

\section{Summary and Discussions}\label{sec_summary}
\setcounter{equation}{0}
We formulated the tunneling rate in the presence of gravity using the WdW canonical quantization and the WKB approximation. Our tunneling formula determines the absolute tunneling rate and there is no need of an ad-hoc normalization. We showed that our formula agrees with the CDL formula except that what is subtracted from the action is determined automatically. We also found that all the four-dimensional equations of motion are reproduced from the Hamiltonian constraint, the momentum constraints and the stationary conditions, which confirms that our formula is consistent with the classical mechanics.

A unique feature of our formulation is that $\alpha^2$ can change its sign locally. It means that, while a region undergoes quantum tunneling, another region may experience the Lorentzian evolution simultaneously. It is a natural consequence of the locality of the WdW equation and is also necessary for the determination of the tunneling rate without knowing all the configuration outside the region of interest.

Once we allow $\alpha^2$ to flip its sign, we can consider tunneling processes that involve both Euclidean and Lorentzian evolution simultaneously. We executed the numerical analysis with an $O(3)$-symmetric Ansatz and found that there are paths that give much higher tunneling rates than that of the CDL bounce: the polychronic tunneling paths. There, we observed that only thin-shell regions have $\alpha^2<0$ and contribute to the tunneling rate.
It is in contrast with the CDL bounce, where the whole region undergoes the Euclidean evolution until it materializes. We also observed that the polychronic tunneling paths exist even in the decoupling regime of gravity and maintain their importance in low energy phenomenology.

It may be possible to generalize our formalism so that $\alpha$ can be complex. In that case, $\alpha^2$ can change its sign without becoming singular, {\it i.e.} $|\alpha| = \infty$, between the Lorentzian and Euclidean regions. If we solve the Einstein-Hamilton-Jacobi equation as an initial value problem,
the solution might not be unique after crossing a point where $|\alpha|=\infty$ because the time derivative is ill-defined there. Although this does not cause a problem in our formulation since we search for a saddle point of Eq.~\eqref{eq_theta_final} directly, a complex $\alpha$ could enable us to solve the initial value problem beyond $|\alpha| = \infty$. These points will be discussed elsewhere.

Let us discuss possible ways to prohibit polychronic tunneling paths from dominating the tunneling rate. What we did in this paper are (i) constructed elements of $\Omega$ in the WKB approximation and found there exist elements that mix Euclidean and Lorentzian evolution, (ii) defined the tunneling wave functionals as those that decay monotonically and connect the false vacuum and escape points toward the true vacuum, and (iii) found complex saddle points that have higher tunneling rates than that with the CDL bounce.
For (i), there is a possibility that the Hamiltonian and the momentum constraints are not the only constraints to define the ``physical'' representation space of quantum gravity. The additional constraints may forbid a part of polychronic tunneling paths.
For (ii), the definition of quantum tunneling beyond the WKB approximation poses many difficulties as explained in Subsection \ref{subsec_tunneling}. Thus, there is a possibility that the ultimate definition of quantum tunneling forbids a part of polychronic tunneling paths.
Finally, for (iii), we have assumed the tunneling rate is given by the decaying factor of the wave functional, which is basically the action integrated along the path. However, there may be an additional suppression factor such as the overall factor of the tunneling rate or a statistical factor. It may suppress the polychronic tunneling by many orders of magnitude and allow the CDL bounce to dominate the tunneling rate.
Finding any of those will have a significant impact on quantum gravity and quantum tunneling and the polychronic tunneling could be a clue for it.

An interpretation of the non-decoupling behavior would be that the PT is a phenomenon like the mixed tunneling in many-body quantum mechanics, {\it e.g.} one of the coupled particles goes through a potential barrier while the others move in real time. Since it is a physical phenomenon, it should be treatable in the correct theory. For a more promising discussion on the PT in the decoupling regime of gravity, however, it is crucial to obtain the same result from a theory without gravity. 
More detailed discussions in this direction are available in \cite{Shoji:2022rke}. Also, our technique is applicable to more realistic situations involving (partially) dynamical background. It was recently proposed that the time-ordered Green's function can be used to formulate vacuum decay in such a situation \cite{Shkerin:2021zbf}. It would be interesting to understand the relation between the technique and ours.

\begin{acknowledgments}
N.O. is supported in part by the Special Postdoctoral Researcher (SPDR) Program at RIKEN, FY2021 Incentive Research Project at RIKEN, and Grant-in-Aid for Scientific Research (KAKENHI) project for FY 2021 (21K20371).
Y.S.~is supported by the I-CORE Program of the Israel Planning Budgeting Committee (grant No. 1937/12). M.Y.~is supported in part by JSPS Grant-in-Aid for Scientific Research Numbers JP18K18764, JP21H01080, JP21H00069. The authors gratefully acknowledge the computational and data resources provided by the Fritz Haber Center for Molecular Dynamics.
\end{acknowledgments}
\appendix
\section{Comparison with the formulation in \cite{Cespedes:2020xpn}}\label{sec_camq}
In this appendix, we show that our formula agrees with that in \cite{Cespedes:2020xpn} when the lapse function does not depend on the spatial position and the signs of $\mathcal K$ and $\mathcal V$ do not flip. We compare the decaying factor of the wave functional before normalizing it since our formulation does not allow such a normalization\footnote{In \cite{Cespedes:2020xpn}, they normalize the wave functional with another wave functional. Then, the total derivative terms of Eq.~\eqref{boundary_action} cancel out and this is why they reproduce the CDL result.}.

In \cite{Cespedes:2020xpn}, they derive the formula given by
\begin{equation}
    \Theta_{\rm CAMQ}^{(0)}=2\int\dd{s}\sqrt{\int\dd[3]{x}\sqrt{h}\mathcal K}\sqrt{\int\dd[3]{x}\sqrt{h}(-\mathcal V)}.\label{eq_camq_formula}
\end{equation}
In their formulation, they assume
\begin{equation}
    \pdv{\Phi^M}{s}=\frac{\gamma^{MN}}{C(s)\sqrt{h}}\fdv{\Theta^{(0)}}{\Phi^N},
\end{equation}
where $C(s)$ is an arbitrary function corresponding to $\alpha^{-1}(s,\bfx)$ in our formulation. Substituting it into Eq.~\eqref{eq_ehj}, we have the classical Hamiltonian constraint,
\begin{equation}
    C^2(s)\mathcal K=-\mathcal V.
\end{equation}
Since they take $C(s)$ as
\begin{equation}
    C(s)=\frac{\sqrt{\int\dd[3]{x}\sqrt{h}(-\mathcal V)}}{\sqrt{\int\dd[3]{x}\sqrt{h}\mathcal K}},
\end{equation}
the Hamiltonian constraint remains unsolved for a general path. To satisfy the Hamiltonian constraint in their formulation, we need to use a classical solution with an Ansatz:
\begin{equation}
    g_{\mu\nu}\dd{x}^{\mu}\dd{x}^{\nu}=\dd{\tau}^2+h_{ij}(\tau,x)\dd{x}^i\dd{x}^j,\label{eq_metric_alpha_i}
\end{equation}
where we rescaled the time variable assuming $C^2(s)<0$ as
\begin{equation}
    \dd{\tau}^2=-\frac{1}{C^2(s)}\dd{s}^2.
\end{equation}

If there is no coordinate transformation from a classical solution to the above metric with keeping the deformation direction, their results cannot be compared with ours. Thus, let us assume there is such a transformation and compare Eq.~\eqref{eq_camq_formula} and Eq.~\eqref{eq_theta_final} using the same classical solution\footnote{Whatever the stationary conditions one obtains, any classical solution should satisfy them since it is what the semi-classical approximation is.}. We obtain
\begin{align}
    \Im\Theta_{\rm CAMQ}^{(0)}&=2\int\dd{s}\sqrt{\int\dd[3]{x}\sqrt{h}\mathcal K}\sqrt{\int\dd[3]{y}\sqrt{h}\mathcal V}\nonumber\\
    &\leq2\int\dd{s}\sqrt{\int\dd[3]{x}\sqrt{h}\abs{\mathcal K}}\sqrt{\int\dd[3]{y}\sqrt{h}\abs{\mathcal V}}\nonumber\\
    &=2\int\dd{s}\sqrt{\int\dd[3]{x}\sqrt{h}(-C^2)^{1/4}\sqrt{\mathcal K\mathcal V}}\sqrt{\int\dd[3]{y}\sqrt{h}(-C^2)^{-1/4}\sqrt{\mathcal V\mathcal K}}\nonumber\\
    &=2\int\dd[3]{x}\sqrt{h}\abs{\sqrt{\mathcal K}\sqrt{\mathcal V}}\nonumber\\
    &=2\Im\int\dd[3]{x}\sqrt{h}\sqrt{\mathcal K}\sqrt{-\mathcal V}\\
    &=\Im\Theta^{(0)},
\end{align}
where the equality holds when $\mathcal V$ and $\mathcal K$ have definite signs.
Thus, our formulation agrees with theirs when there is the transformation and there is no (simultaneous) flip of the signs of $\mathcal V$ and $\mathcal K$.

\section{Importance of choosing appropriate coordinates}\label{apx_foliation}
Although our result agrees with \cite{Cespedes:2020xpn}, our formulation is indispensable for the description of bubble nucleation due to the following reason.

As we have seen in Sec.~\ref{sec_cdl}, the CDL bounce can be transformed into the static coordinates as
\begin{align}
    \phi&=\phi(\tau,r),\\
    g_{\mu\nu}\dd{x^\mu} \dd{x^\nu}&=X^2(\tau,r)\dd{\tau}^2+A^2(\tau,r)\dd{r}^2+r^2(\dd{\theta}^2+\sin^2\theta \dd{\phi}^2).
\end{align}
Here, we take $\tau$ so that $\tau=0$ corresponds to the turning point and $\tau=\tau_i$ corresponds to the initial state. 
As explained in Appendix \ref{sec_camq}, the formulation of \cite{Cespedes:2020xpn} requires a solution in the form of \eqref{eq_metric_alpha_i}.
Thus, we consider the transformation from $(\tau,r)$ to $(t(\tau,r),x(\tau,r))$. To make $g_{tt}=1$ and $g_{tx}=0$, we have
\begin{align}
    X^2\qty(\pdv{x}{r})^2+A^2\qty(\pdv{x}{\tau})^2&=\qty(\pdv{t}{\tau}\pdv{x}{r}-\pdv{t}{r}\pdv{x}{\tau})^2,\\
    X^2\pdv{x}{r}\pdv{t}{r}+A^2\pdv{x}{\tau}\pdv{t}{\tau}&=0.
\end{align}
Since we want to discuss the rate to have the final state of the CDL bounce, we solve the above differential equations with the following boundary conditions:
\begin{equation}
    t(0,r)=0,~x(0,r)=r,~t(\tau,0)=\tau,~x(\tau,0)=0.
\end{equation}

It is instructive to see what happens for the path where the fields stay at the false vacuum,
\begin{align}
    \phi=\phi_{\rm FV},~X=(1-\bar H^2r^2)^{1/2},~A=(1-\bar H^2r^2)^{-1/2},
\end{align}
where $\bar H=\sqrt{2\kappa V(\phi_{\rm FV})/3}$ is a Hubble constant.
Apparently, in the coordinates of $(\tau,r)$, we have $\mathcal V=\mathcal K=0$ and the wave functional does not decay.

\begin{figure}
    \centering
    \includegraphics[width=0.6\linewidth]{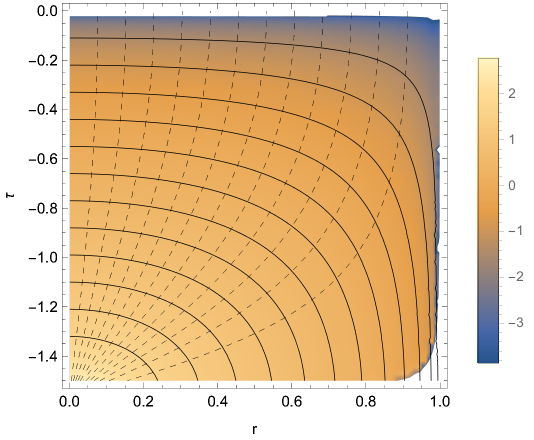}
    \caption{The new coordinates with $\bar H=1$ and the value of $\kappa\mathcal K=\kappa\mathcal V$. The solid lines are the contours of $t$ and the dashed ones are those of $x$. $\tau=t$ at $r=0$ and $x=r$ at $\tau=0$. The color map shows $\log_{10}(-\kappa\mathcal K)$. The behavior at the bottom right corner is a numerical artifact.}
    \label{fig_coord}
\end{figure}

After the transformation, however, we obtain a different result. The new axes are shown in Fig.~\ref{fig_coord}, where we take $\bar H=1$. The solid lines are contours of $t$ and the dashed ones are those of $x$. The coordinates collapse at $\tau=t=-3/2$. Let us consider a transition from an initial state at $t=t_i>-3/2$ to the final state at $t=0$. The value of $\kappa\mathcal V=\kappa\mathcal K$ is indicated with the color map. As we can see, $\mathcal K\neq0 (\alpha^2<0)$ over the region and thus the wave functional decays as $t$ evolves.
More importantly, it also means that the initial state is inevitably virtual. Thus, the result cannot even be interpreted as a tunneling rate.

If we reverse $t$, this process can be interpreted as follows; when the initial state at $t=0$ evolves in a way that the spatial curvature increases with a constant $\phi$, the final state is virtual and the wave functional decays toward this direction. Therefore, it just describes the evanescent wave.

This example clearly shows that it is important to choose the coordinate system appropriately to discuss the tunneling rate in our formulation. In particular, we should use the coordinates that do not evolve at the false vacuum to discuss the CDL bubble nucleation, which is why we adopt the static coordinates.

\section{Shift functions}\label{apx_shift}
We present the formulas with shift functions. All the indices are raised and lowered with $h^{ij}$ and $h_{ij}$. For the discussion of integrability, see \cite{Shoji:2022rke}.

At the leading order in $\hbar$, the momentum constraints, Eq.~\eqref{eq_mom_const}, are given by
\begin{equation}
    0=\beta_i(s,\bfx)\qty[(\partial^i\phi)\fdv{\Theta^{(0)}(s)}{\phi(\bfx)}-2\sqrt{h}\nabla_j\frac{1}{\sqrt{h}}\fdv{\Theta^{(0)}(s)}{h_{ij}(\bfx)}],
\end{equation}
where we multiplied arbitrary functions, $\beta_i(s,\bfx)$. It can be rewritten as
\begin{equation}
    w^M\fdv{\Theta^{(0)}(s)}{\Phi^M(\bfx)}=2\partial_i\beta_j\fdv{\Theta^{(0)}(s)}{h_{ij}(\bfx)},
\end{equation}
where
\begin{align}
    w^\phi&=\beta^i\partial_i\phi,\\
    w^{(ij)}&=\nabla_i\beta_j+\nabla_j\beta_i.
\end{align}
Adding this to Eq.~\eqref{eq_ehj}, we have
\begin{equation}
    (v^M+w^M)\fdv{\Theta^{(0)}(s)}{\Phi^M(\bfx)}=-2\alpha\sqrt{h}\mathcal V+2\partial_i\beta_j\fdv{\Theta^{(0)}(s)}{h_{ij}(\bfx)}.
\end{equation}
To construct a solution along the path, we need
\begin{equation}
    \pdv{\Phi^M(s,\bfx)}{s}=v^M+w^M.
\end{equation}
which can be seen as a constraint on $\Theta^{(0)}$:
\begin{equation}
    \fdv{\Theta^{(0)}(s)}{\Phi^M(\bfx)}=\frac{\sqrt{h}}{\alpha}\gamma_{MN}\qty(\pdv{\Phi^N(s,\bfx)}{s}-w^N).
\end{equation}
It can be solved as
\begin{align}
    \Theta^{(0)}(s_f)-\Theta^{(0)}(s_i)&=\int\dd[4]{x}\frac{2\sqrt{h}}{\alpha}\mathcal K+\int\dd[4]{x}\frac{\sqrt{h}}{\alpha}\gamma_{MN}w^M\qty(\pdv{\Phi^N(\bfx)}{s}-w^N)\nonumber\\
    &=\int\dd[4]{x}\frac{2\sqrt{h}}{\alpha}\mathcal K+\int\dd[4]{x}\beta^iC_i^{\mathcal M}+\int\dd[4]{x}\partial_i\sqrt{h}\mathcal J^i,\label{eq_shift_anzats}
\end{align}
where we redefined
\begin{align}
    \mathcal K&=\frac12\gamma_{MN}\qty(\pdv{\Phi^M(s,\bfx)}{s}-w^M)\qty(\pdv{\Phi^N(s,\bfx)}{s}-w^N)\nonumber\\
    &=\frac12(\partial_s\phi-\beta^i\partial_i\phi)^2+\frac{G^{ijkl}}{2}(\partial_sh_{ij}-2\nabla_i\beta_j)(\partial_sh_{kl}-2\nabla_k\beta_l),\label{eq_k_with_beta}
\end{align}
and
\begin{align}
    C^{\mathcal M}_i&=\frac{\sqrt{h}}{\alpha}(\partial_i\phi)(\partial_s\phi-\beta^k\partial_k\phi)-h_{ip}\frac{\sqrt{h}}{2\kappa}\nabla_j\qty[\frac{G^{pjkl}}{\alpha}(\partial_sh_{kl}-2\nabla_k\beta_l)],\\
    \mathcal J^i&=\frac{\beta_j}{2\kappa\alpha}G^{ijkl}(\partial_sh_{kl}-2\nabla_k\beta_l).\label{eq_def_j}
\end{align}

The momentum constraints give the classical momentum constraints, $C^{\mathcal M}_i=0$, and
the Hamiltonian constraint gives the classical Hamiltonian constraint, $\alpha^2=-\mathcal K/\mathcal V$. After applying the classical Hamiltonian constraint and the classical momentum constraints, the stationary conditions give the $(ij)$-elements of Einstein equations and the equations of motion for the scalar field.

The total derivative terms appearing in the Gauss-Codacci equation are modified as
\begin{align}
    \mathcal X^s&=-\frac{h^{kl}}{\alpha^2}\qty(\partial_sh_{kl}-2\nabla_k\beta_l),\\
    \mathcal X^i&=\frac{1}{\alpha^2}\qty[\partial^i\alpha^2+\beta^ih^{kl}(\partial_sh_{kl}-2\nabla_k\beta_l)].
\end{align}
Using these and Eq.~\eqref{eq_shift_anzats}, we find that $\Theta^{(0)}$ can be expressed as
\begin{equation}
    \Theta^{(0)}=\frac{S}{2}+\int\dd[4]{x}\partial_\mu\mathcal H_{\rm bdy}^\mu.
\end{equation}

\section{Existence of a saddle point of $\Re\Theta^{(0)}$ for a fixed $\Im\Theta^{(0)}$}
\label{apx_re_saddle}
Here, we show that we do not need to find a saddle point of $\Re\Theta^{(0)}$ for the evaluation of a tunneling rate if that of $\Im\Theta^{(0)}$ is found. We assume $\mathcal K\geq0$ with the branch cuts of Eq.~\eqref{eq_alpha_def} in the following, but we can show it also for the other cases.

Let us assume there is a path $\bar\Phi_0^M(s,\bfx)$ that satisfies
\begin{align}
    \eval{\int\dd[4]{x}w^M(s,\bfx)\fdv{\Im\Theta^{(0)}}{\Phi^M(s,\bfx)}}_{\Phi^M=\bar\Phi_0^M}=0,
\end{align}
for all the functions\footnote{To simplify the discussion, we ignore the classical momentum constraints, which $\Phi^M$ needs to satisfy. One can make the discussion rigid by constraining $w^M(s,\bfx)$ (and also $z^M(s,\bfx)$) so that $\bar\Phi_0^M+\epsilon w^M$ satisfies the linearized classical momentum constraints around $\bar\Phi_0^M$.}, $w^M(s,\bfx)$, satisfying $w^M(s_i,\bfx)=w^M(s_f,\bfx)=0$, {\it i.e.} $\bar\Phi^M_0$ is a saddle point of $\Im\Theta^{(0)}$ with fixed initial and final states.

We assume there are flat directions of $\Im\Theta^{(0)}$;
\begin{align}
    \eval{\int\dd[4]{x}z^M(s,\bfx)\frac{\delta^2\Im\Theta^{(0)}}{\delta\Phi^M(s,\bfx)\delta\Phi^N(s',\bfx')}}_{\Phi^M=\bar\Phi_0^M}=0,
\end{align}
for $z^M(s,\bfx)\in \mathcal M_{\rm flat}$. Here, $\mathcal M_{\rm flat}$ is the set of flat directions that satisfy $z^M(s_i,\bfx)=z^M(s_f,\bfx)=0$ for all $\bfx$ and $M$.

If $\bar\Phi_0^M(s,\bfx)$ is not a saddle point of $\Re\Theta^{(0)}$ for a fixed\footnote{The paths with different $\Im\Theta^{(0)}$ do not cancel with each other.} $\Im\Theta^{(0)}$, there exists $z_0^M(s,\bfx)\in \mathcal M_{\rm flat}$ such that
\begin{align}
    \eval{\int\dd[4]{x}z_0^M(s,\bfx)\fdv{\Re\Theta^{(0)}}{\Phi^M(s,\bfx)}}_{\Phi^M=\bar\Phi_0^M}<0.
\end{align}
With an infinitesimal parameter, $\epsilon>0$, we define
\begin{equation}
    \bar\Phi^M_1(s,\bfx)= \bar\Phi^M_0(s,\bfx)+\epsilon z_0^M(s,\bfx).
\end{equation}
Then,
\begin{align}
    \Re\Theta^{(0)}\qty[\bar\Phi_1^M]&<\Re\Theta^{(0)}\qty[\bar\Phi_0^M].
\end{align}
Since $\Re\Theta^{(0)}>0$, for an arbitrary number, $\delta>0$, there exists an integer $\mathcal N>0$ and $\Re\Theta^{(0)}_\infty>0$ such that
\begin{equation}
    \abs{\Re\Theta^{(0)}\qty[\bar\Phi_n^M]-\Re\Theta^{(0)}_\infty}<\delta,
\end{equation}
for all $n>\mathcal N$. Here, $\bar\Phi_n^M(s,\bfx)$ is the path after repeating the above procedure for $n$-times. We also have
\begin{align}
    \eval{\fdv{\Im\Theta^{(0)}}{\Phi^M(s,\bfx)}}_{\Phi^M=\bar\Phi_n^M}&=\order{n\epsilon^2}.
\end{align}
Taking the limit of $\epsilon\to0$ with keeping $\epsilon n$ constant, we can make $\delta\Im\Theta^{(0)}/\delta\Phi^M$ arbitrarily small for a given $\delta$.

If $\lim_{n\to\infty}\bar\Phi_n^M$ exists, it is a saddle point of $\Theta^{(0)}$ having the same $\Im\Theta^{(0)}$. It is also possible that there is an approximately flat direction and the limit does not exist. Since there is no cancellation of phases for $\delta\ll2\pi$, the tunneling rate diverges proportionally to the volume of the flat direction. In such a case, what we can calculate is only the tunneling rate divided by the volume, and its exponent is the same as the one without the flat direction. Thus, to obtain the exponent of the tunneling rate, it is sufficient to find a minimum of $\Im\Theta^{(0)}$.
\section{Numerical optimization of $\Re\Theta^{(0)}$}\label{apx_osc}
Here, we show the results with optimizing (minimizing) both $\Re\Theta^{(0)}$ and $\Im\Theta^{(0)}$. We define
\begin{align}
    \ln\varepsilon[\phi(s,r)]&=-16\pi\int_{0}^{1}\dd{s}\int_0^{r_{\max}}J(s,r),
\end{align}
where
\begin{align}
    J(s,r)&=r^2e^{\eta/2}\sqrt{\mathcal K}\Im\qty(\sqrt{\mathcal V}).
\end{align}

\begin{figure}[t]
  \begin{minipage}{0.49\linewidth}
    \includegraphics[width=\linewidth]{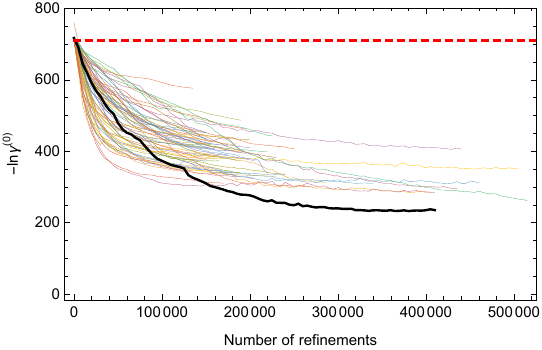}
  \end{minipage}
  \begin{minipage}{0.49\linewidth}
    \includegraphics[width=\linewidth]{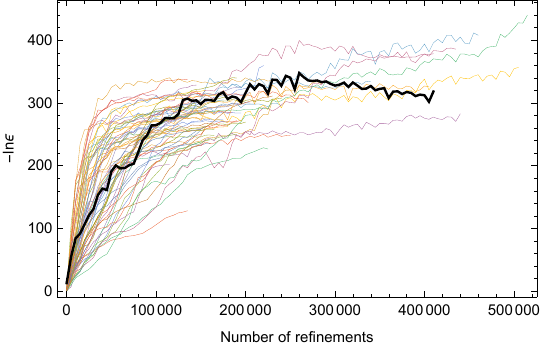}
  \end{minipage}
  \begin{minipage}{0.49\linewidth}
    \includegraphics[width=\linewidth]{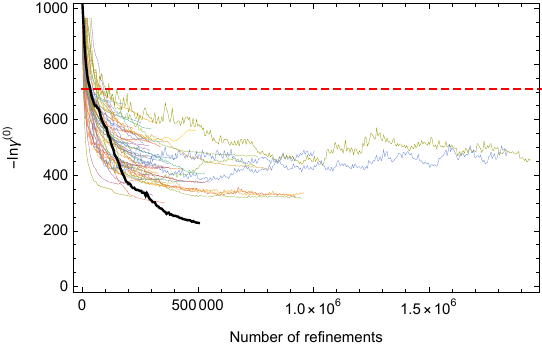}
  \end{minipage}
  \begin{minipage}{0.49\linewidth}
    \includegraphics[width=\linewidth]{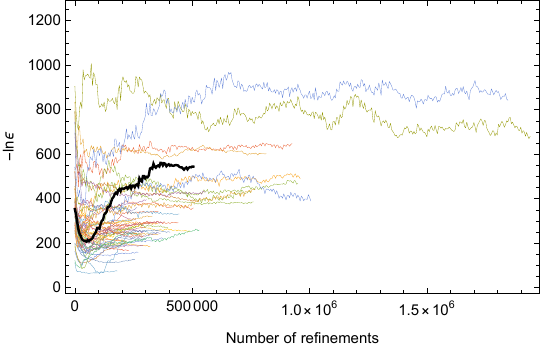}
  \end{minipage}
\caption{Optimization histories with $(\kappa,k,V_0)=(0.5,0.3,0)$. The left panels show $\ln\gamma^{(0)}$ and the right ones show $\ln\varepsilon$. The top panels are for the CDL initial path and the bottom ones are for the Gaussian initial path. The red dashed line indicates the CDL tunneling rate. The lines with the same length and color correspond to the same history. The thick black lines show the history for the highest tunneling rate.}
\label{fig_opt_histry_osc}
\end{figure}
\begin{figure}[t]
  \begin{minipage}{0.49\linewidth}
    \includegraphics[width=\linewidth]{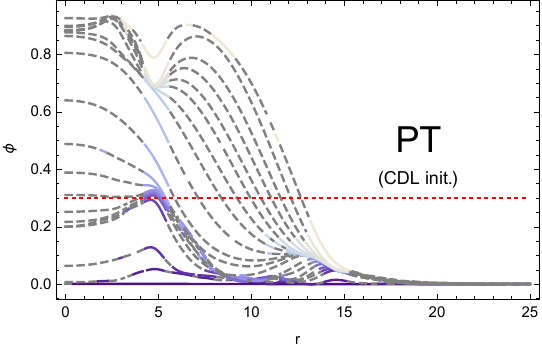}
  \end{minipage}
  \begin{minipage}{0.49\linewidth}
    \includegraphics[width=\linewidth]{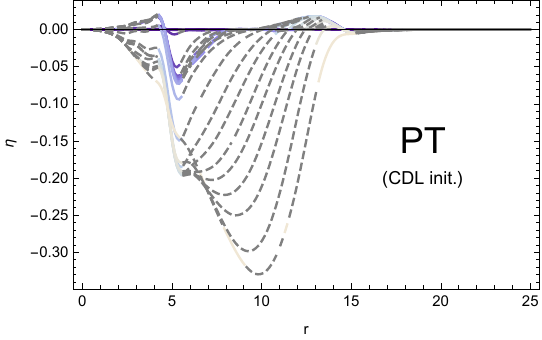}
  \end{minipage}
  \begin{minipage}{0.49\linewidth}
    \includegraphics[width=\linewidth]{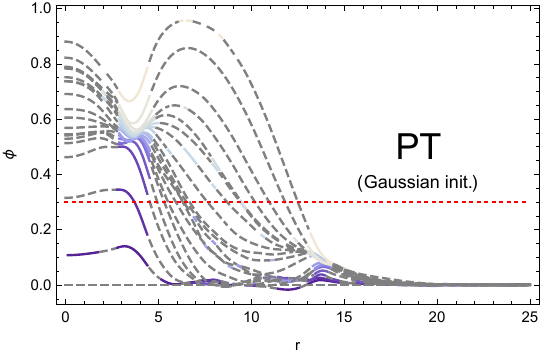}
  \end{minipage}
  \begin{minipage}{0.49\linewidth}
    \includegraphics[width=\linewidth]{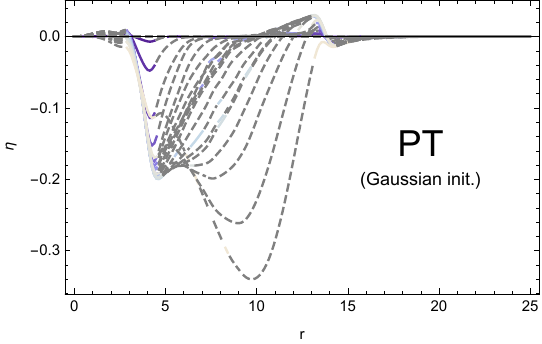}
  \end{minipage}
\caption{The same figures as in Fig.~\ref{fig_profile_1} but with $\Re\Theta^{(0)}$ optimized. The parameters are $(\kappa,k,V_0)=(0.5,0.3,0)$. The top panels are for the CDL initial path and the bottom panels are for the Gaussian initial path.}
\label{fig_profile_apx}
\end{figure}
\begin{figure}[t]
  \begin{minipage}{0.49\linewidth}
    \includegraphics[width=\linewidth]{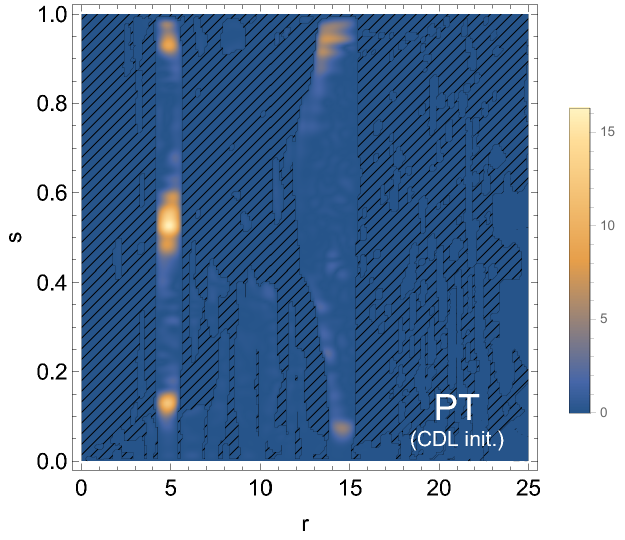}
  \end{minipage}
  \begin{minipage}{0.49\linewidth}
    \includegraphics[width=\linewidth]{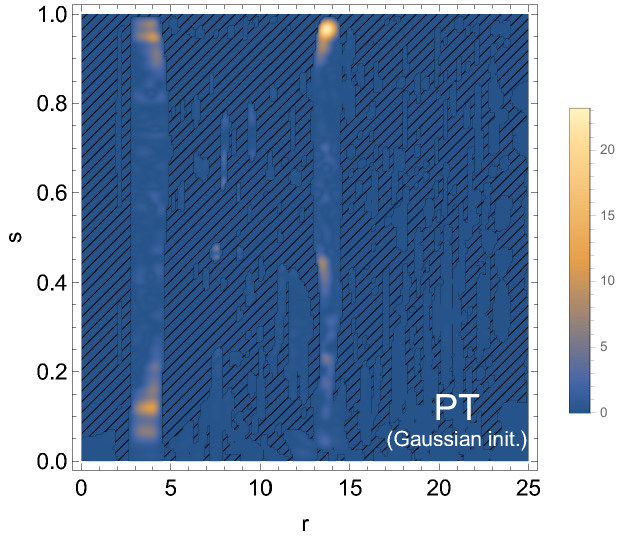}
  \end{minipage}
\caption{The same figures as in Fig.~\ref{fig_integrand_1} but with $\Re\Theta^{(0)}$ optimized. The parameters are $(\kappa,k,V_0)=(0.5,0.3,0)$. The left panel is for the CDL initial path and the right panel is for the Gaussian initial path.}
\label{fig_integrand_apx}
\end{figure}

In this appendix, we take $\kappa=0.5$, $k=0.3$ and $V_0=0$. We repeat the optimization of $\ln\gamma^{(0)}$ for 200 refinements and optimization of $\ln\varepsilon$ for 100 refinements in turn. The optimization history is shown in Fig.~\ref{fig_opt_histry_osc} and the paths with the highest tunneling rate are shown in Figs.~\ref{fig_profile_apx} and \ref{fig_integrand_apx}. The left panels show $\ln\gamma^{(0)}$ and the right ones show $\ln\varepsilon$. The top panels are for the CDL initial path and the bottom ones are for the Gaussian initial path. 

Since the maximization of $\ln\gamma^{(0)}$ prefers less $\alpha^2<0$ region and that of $\ln\varepsilon$ prefers less $\alpha^2>0$ region, the lines fluctuate inevitably. In fact, the lines with highest numbers of refinements exhibit random walks due to the accidental balance between them. In addition, the maximization of $\ln\varepsilon$ stints the evolution after the tunneling and optimizes the final state severely, which increases the probability to be caught by a local maximum.
Even so, for the path with the highest tunneling rate, both of $\ln\gamma^{(0)}$ and $\ln\varepsilon$ seem to approach their local maxima and the change of $\ln\varepsilon$ is $\order{2\pi}$ during the last $10^5$ refinements.
We get the highest tunneling rate of $\ln\gamma^{(0)}\simeq-232$ for the CDL initial path and $\ln\gamma^{(0)}\simeq-228$ for the Gaussian initial path, which are in agreement with those without optimizing $\ln\varepsilon$. 

Since the path we obtained in this appendix is an approximate saddle point of $\Theta^{(0)}$, we can discuss the field evolution of the regions with $\alpha^2>0$ as well. From the top left panel of Fig.~\ref{fig_profile_apx}, we have the following interpretation of the path; (i) the regions around $r\sim5$ and $r\sim15$ precede the deformation and enter the tunnel, (ii) they pull the regions nearby until the innermost region goes over the potential top ($\phi=0.3$), (iii) the innermost region and the region around $r\sim10$ roll down the potential and overtake the tunneling regions, (iv) the tunneling regions are pulled by the rolling regions and escape from the tunnel. The bottom left panel of Fig.~\ref{fig_profile_apx} has a similar behavior but the tunneling region around $r\sim5$ pulls the field more strongly.

\section{Numerical method}\label{apx_numerical}
Here, we describe the numerical method to search for the path with the highest tunneling rate. We assume the potential of Eq.~\eqref{eq_potential}, where the false vacuum is at $\phi=0$ and the true vacuum is at $\phi=1$.
The basic strategy is to execute the Monte Carlo optimization on a 2-dimensional lattice defined on $(s,r)\in[0,1]\otimes[0,r_{\max}]$.
We adopt the following basis functions:
\begin{equation}
    \delta\phi(s,r)=c_he^{-(r^2-\bar r^2)^2/(2w_r\bar r)^2}e^{-(s-\bar s)^2/w_s^2}\sin\frac{\pi}{2}s.
\end{equation}
Here, we chose the function of $r$ so that it becomes the Gaussian function for $\bar r\gg w_r$ and $\partial_r\delta\phi$ becomes zero at $r=0$, while that of $s$ so that $\delta\phi$ becomes zero at $s=0$. The range of the parameters is
\begin{align}
    &\log_{10}|c_h|\in[-4,-1],~\bar s\in[0,1],~w_s\in[3/N_s,0.2],\nonumber\\
    &\bar r\in[0.1r_{\max},0.9r_{\max}],~w_r\in[3r_{\max}/N_r,0.2r_{\max}],
\end{align}
where $N_s$ is the lattice size for $s$ and $N_r$ is that for $r$. These parameters are chosen randomly from the uniform distribution on the logarithmic scale for $c_h$ and on the linear scale for the others.
Here, we chose the widths so that they are at least three times larger than the lattice spacing. We also avoided $\bar r\sim0$ to keep $\partial_r\delta\phi=0$ at $r=0$, and $\bar r\sim r_{\max}$ to keep the solution well inside the lattice. Notice that we do not fix the final state in this calculation.

We first search for the final state that has $\alpha^2>0$.
Taking the last few slices in $1-\delta s<s<1$, we optimize $\phi(s,r)$ so that $U=0$, where
\begin{equation}
    U=\int_{1-\delta s}^1\dd{s}\int_0^{r_{\max}}\dd{r} r^2\max[\mathcal V(s,r),0].
\end{equation}
Once such $\phi(s,r)$ is found, we maximize $\ln\gamma^{(0)}$ and $\ln\varepsilon$ keeping $U=0$.

The flow of the program is as follows.
\begin{enumerate}
    \item Set an initial guess to $\phi$.
    \item Loop the following.
    \begin{enumerate}
        \item Generate $\delta\phi(s,r)$.
        \item Calculate $\eta$ and $U$ with $\phi+\delta\phi$.
        \item If $U$ decreases, set $\phi+\delta\phi$ to $\phi$.
        \item If $U=0$, end the loop.
    \end{enumerate}
    \item Loop the following.
    \begin{enumerate}
    \item Calculate $\ln\gamma^{(0)}$ with $\phi$.
    \item Loop the following for 200 refinements.
    \begin{enumerate}
        \item Generate $\delta\phi(s,r)$.
        \item Calculate $\eta$ and $U$ with $\phi+\delta\phi$.
        \item If $U=0$, calculate $\ln\gamma^{(0)}$ with $\phi+\delta\phi$.
        \item If $\ln\gamma^{(0)}$ increases, set $\phi+\delta\phi$ to $\phi$.
    \end{enumerate}
    \item Calculate $\ln\varepsilon$ with $\phi$.
    \item Loop the following for 100 refinements.
    \begin{enumerate}
        \item Generate $\delta\phi(s,r)$.
        \item Calculate $\eta$ and $U$ with $\phi+\delta\phi$.
        \item If $U=0$, calculate $\ln\varepsilon$ with $\phi+\delta\phi$.
        \item If $\ln\varepsilon$ increases, set $\phi+\delta\phi$ to $\phi$.
    \end{enumerate}
\end{enumerate}
\end{enumerate}
We can omit 3(c) and 3(d) if we optimize only $\ln\gamma^{(0)}$. If we optimize both $\ln\gamma^{(0)}$ and $\ln\varepsilon$, it is important to do similar numbers of refinements for both of them since otherwise only one of them is optimized preferentially due to the statistics.

\section{Other paths with high tunneling rates}
\label{apx_others}
\begin{figure}[h]
    \includegraphics[width=\linewidth]{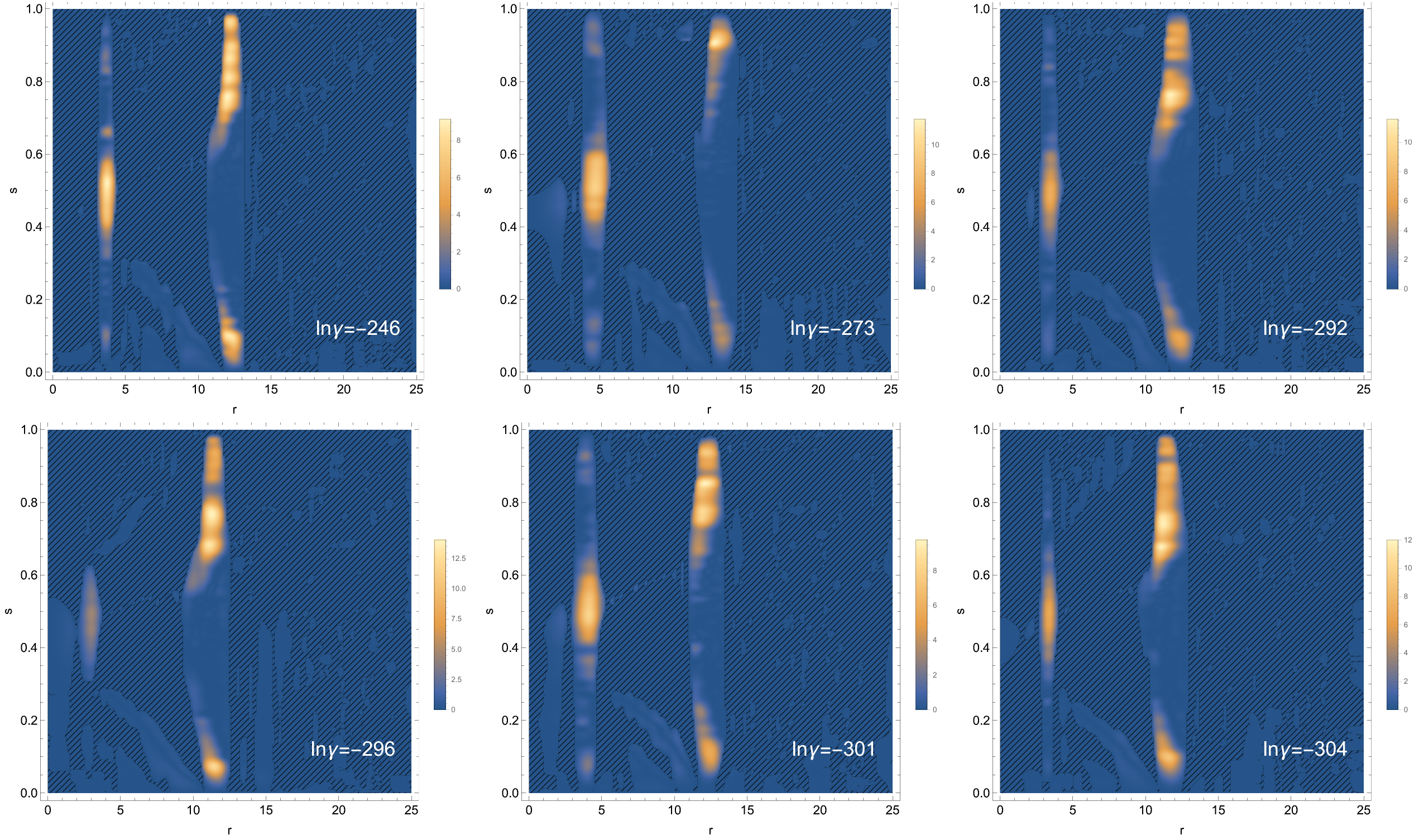}
\caption{The integrand, $I(s,r)$, for other paths with high tunneling rates. The parameters are $(\kappa,k,V_0)=(0.5,0.3,0)$ and the initial path is CDL.}
\label{fig_grid1}
\end{figure}

\begin{figure}[h]
    \includegraphics[width=\linewidth]{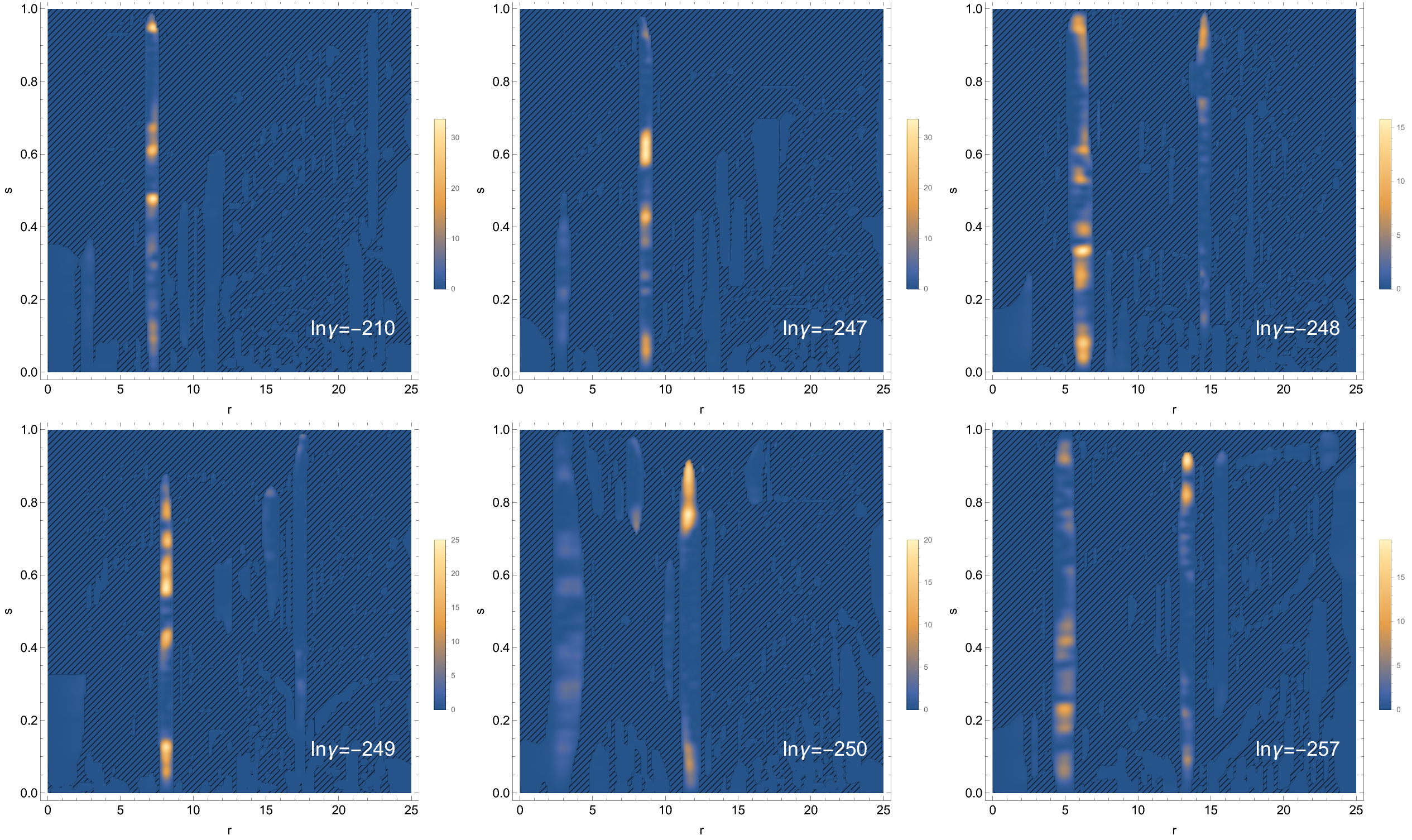}
\caption{The integrand, $I(s,r)$, for other paths with high tunneling rates. The parameters are $(\kappa,k,V_0)=(0.5,0.3,0)$ and the initial path is Gaussian.}
\label{fig_grid2}
\end{figure}

In Figs.~\ref{fig_grid1} and \ref{fig_grid2}, we show $I(s,r)$ for other paths with high tunneling rates. Notice that the final states are different for different paths. The parameters are those used in Fig.~\ref{fig_integrand_1}, {\it i.e.} $(\kappa,k,V_0)=(0.5,0.3,0)$. Fig.~\ref{fig_grid1} is with the CDL initial path and Fig.~\ref{fig_grid2} is with the Gaussian initial path.

With the CDL initial path, the tunneling regions are similar; there are brighter regions around $r\sim10-15$ and $r\sim2-5$.
On the other hand, with the Gaussian initial paths, we find one or two brighter regions. In all these paths, the tunneling regions appear vertically unlike in the CDL bounce.
\bibliographystyle{apsrev4-1}
\bibliography{wdw}
\end{document}